\documentclass[twocolumn,nobibnotes, aps, prb, superscriptaddress,floatfix]{revtex4-2}

\usepackage[normalem]{ulem} 	
\usepackage[usenames,dvipsnames]{color}
\usepackage{upgreek}
\usepackage{float}
\usepackage[nointegrals]{wasysym}
\usepackage{helvet}
\usepackage{hyperref}
\usepackage{xcolor}
\usepackage[english]{babel}
\usepackage{graphicx}
\usepackage[export]{adjustbox}
\usepackage{amssymb,amsfonts,amsmath} 
\usepackage{physics}

\definecolor{pink2}{RGB}{254, 1, 154}

\newcommand{\eq}[1]{\begin{align}#1\end{align}}

\begin{document}

\title{Measurements on an Anderson Chain}
\author{Paul Pöpperl}
	\affiliation{\mbox{Institut für Theorie der Kondensierten Materie, Karlsruhe Institute of Technology, 76128 Karlsruhe, Germany}}
	\affiliation{\mbox{Institute for Quantum Materials and Technologies, Karlsruhe Institute of Technology, 76021 Karlsruhe, Germany}}
\author{Igor~V.~Gornyi}
\affiliation{\mbox{Institut für Theorie der Kondensierten Materie, Karlsruhe Institute of Technology, 76128 Karlsruhe, Germany}}
\affiliation{\mbox{Institute for Quantum Materials and Technologies, Karlsruhe Institute of Technology, 76021 Karlsruhe, Germany}}
\affiliation{A. F. Ioffe Institute, 194021 St. Petersburg, Russia}
\author{Yuval Gefen}
\affiliation{\mbox{Department of Condensed Matter Physics, Weizmann Institute of Science, 7610001 Rehovot, Israel}}
\date{\today}

\begin{abstract}
    We study the dynamics of a monitored single particle in a one-dimensional, Anderson-localized system. The time evolution is governed by Hamiltonian dynamics for fixed time intervals, interrupted by local, projective measurements. The competition between disorder-induced localization and measurement-induced jumps leads to interesting behaviour of readout-averaged quantities. We find that measurements at random positions delocalize the average position, similar to a classical random walk. Along each quantum trajectory, the particle remains localized, however with a modified localization length. In contrast to measurement-induced delocalization, controlled measurements can be used to introduce transport in the system and localize the particle at a chosen site. In this sense, the measurements provide a controlled environment for the particle.
\end{abstract}

\maketitle

\section{Introduction}
The interplay between unitary time evolution and measurements of varying strength and frequency in different systems is a recently very active subject of research \cite{Fisher-PhysRevB.98.205136,measured_quantum_circuits_1,measurement_entanglement_transition_quantum_circuits,Chan-PhysRevB.99.224307,Altman-PhysRevB.101.104301,Romito-PhysRevB.100.064204,quantum_circuits_review,entanglement_transition_monitored_free_fermion_chain, measured_dirac_fermions, DGG_measured_chain, entanglement_transitions_with_free_fermions,entanglement_continuously_mintored_chain,entangelement_transitions_quasiparticles,
dynamics_of_measured_mb_chaotic_systems,interplay_measurement_decoherence_free_hamiltonian_evolution,continuous_measurement_free_bosons,continuously_measured_free_fermion_gas}. Examples for emergent dynamical effects are transitions in the asymptotics of the entanglement entropy in quantum circuit models \cite{quantum_circuits_review} and Hamiltonian systems \cite{entanglement_transition_monitored_free_fermion_chain, measured_dirac_fermions, DGG_measured_chain, entanglement_transitions_with_free_fermions,entanglement_continuously_mintored_chain,entangelement_transitions_quasiparticles}, as well as statistical properties of measurement outcomes in quantum lattices~\cite{quantum_time_of_arrival, projective_measurements_tb}. Similar observations are made in related systems with noise or dissipation instead of measurements \cite{entanglement_production_dissipative_impurity, fermionic_chain_with_dissipative_defects,stochastic_resistors}.

If disorder is introduced into a one-dimensional non-interacting chain, the system becomes Anderson-localized~\cite{paper:absence_of_diffusion,paper:localization_in_two_d,Evers-RevModPhys}. All eigenstates decay exponentially, and transport from one end of the chain to the other is exponentially suppressed with the size of the system. Because of this special property of the eigenstates, it is natural to ask about measurements in this context. In particular, one may wonder whether the introduction of measurements leads to dephasing, destroying localization and establishing transport. A related issue of noise-induced dynamics in a localized system is discussed in Ref.~\cite{gopalakrishnan_knap_noise_and_diffusion}.
Recently, the effect of measurements on the many-body localized phase has been studied~\cite{mbl_measurements}. Considering a single-particle model has the benefit of being more tractable numerically, and intuitively comprehensible. At the same time, we find interesting physics in the simple model. 

In the present work, we consider local projective measurements of the site occupation. If such a measurement occurs within the localization length of a localized particle, the particle is often detected at the measured site. In these cases, the center of the wave function shifts to the measured site as a consequence of the projection, and the wave packet starts spreading around this site. This spreading may be limited by the localization volume, by the next detection of the particle, or by some putative ``dephasing'' mechanism introduced by the measurement backaction. Repeated measurements are therefore expected to induce transport in the system. With this idea in mind, we follow two general directions: On the one hand, we choose measurement locations at random and investigate the consequent dynamics. On the other hand, we try to manipulate the state of the system in a controlled way, by designing ``measurement protocols" that aim for spatial steering of the particle. The engineering and manipulation of quantum states through measurements has been explored in Refs.~\cite{steering_rcgg,Herasymenko,measurements_restructure_critical_states,Klich-PhysRevX.12.031031}.

Based on the fact that a projective measurement collapses the particle's wave function to the measured site when occupation one is measured (``click outcome"), there are different intuitive expectations that one may have for the average behaviour of the particle position. On the one hand, in the limit of very frequent measurements, the particle is confined to its initial site as a consequence of the quantum Zeno effect \cite{Misra1977, Peres1980, Chaudhry2016, Snizhko2020}. On the other hand, if the time between two measurements is sufficiently large for the time evolution to spread the wave packet to its exponential envelope, diffusion would be a natural expectation for the spread of an ensemble of particle positions. The latter case of large time intervals between two measurements is investigated in this work.

While diffusion may first come to mind, thinking about a click-driven ``classical random walk" of the localized wave function, the distribution of localization lengths may spoil this behaviour. In fact, the site with the largest probability to host a click event is typically the center of the wave function, where the previous click has occurred. Depending on the probability of long successions of click events on the same site, ``waiting" of the particle in rare regions of small localization lengths may potentially lead to subdiffusion. Indeed, it was shown in Ref.~\cite{gopalakrishnan_knap_noise_and_diffusion} that rare regions can provide bottlenecks for the dynamics in the related case of a disordered system with temporal noise, leading to subdiffusion on intermediate, parametrically large, time scales.

Another issue is related to the impact of ``no-click events'' on the wave function. A no-click event, measuring zero occupation on the site, produces a ``hole'' at the measured position, where the particle is then known to not be. The total probability to find the particle is consequently redistributed by the normalization among all other sites. A priori, it is not clear whether this procedure of making a hole and renormalizing the wave function favours localization or delocalization. Localization may be favoured, because no-click events are more likely to occur in the tail of the wave function, removing weight from the tail and shifting it towards the center via normalization, such that the probability of a subsequent click event within the localization volume would increase. Delocalization may be favoured, because holes close to the center have a larger impact on the wave function, as a larger portion of the wave function is redistributed---also into the tails. Overall, the no-click events could then enhance the probability of large localization lengths such that long jumps due to click events become more likely.

In this paper, we formalize and investigate these questions about the fate of localization of the wave function, as well as transport in the ensemble of particle positions, under sequences of repeated projective measurements. We follow individual quantum trajectories (sequences of measurement outcomes for given sequences of measurement positions in a fixed disorder realization) to acquire statistics that allows us to obtain averaged observables.

In the case of random measurement locations, our key observation is that while the particle position is randomized over the entire system, the wave packet typically remains exponentially localized around its center site. Considering averages over measurement outcomes, measurement locations, as well as over disorder realizations, we investigate the spread of the ensemble of wave-packet centers throughout the system and modifications to the localization lengths due to the measurements. Supported by a connection to a classical random walk, we argue that the particle trajectories of different random realizations spread diffusively in the long-time limit.  The idea of measurement-induced random walks has been addressed in Ref.~\cite{didi2021measurement}. In Ref. \cite{hugo_random_walks}, the authors use a mapping to a classical random walk to model dynamical properties of quantum systems subject to measurements and disorder.

Regarding steering, we analyze and compare different kinds of protocols, either using the detector readout only at a designated target site, or at every measurement location. Having access to all readouts allows us to induce ballistic transport. 
In this sense, both types of steering, passive~\cite{steering_rcgg} and active~\cite{Herasymenko}, are considered. Diffusion due to random measurements allows for polynomial steering times, even if only reading out the target site.

The paper is structured as follows. We introduce the time evolution protocol for random measurements in Sec. \ref{sec:hamiltonian_and_protocol}. In Sec. \ref{sec:particle_trajectories}, we describe the resulting dynamics qualitatively, motivating the discussion of particle trajectories and corresponding observables. With these observables, we investigate ``delocalization" due to measurements in Sec. \ref{sec:measurement_delocalization}. We present the numerical results on steering by non-random measurements in Sec. \ref{sec:steering}. Finally, in Sec. \ref{sec:random_walk}, the relation to a classical random walk is formalized, and, based on that, our numerical results are further discussed. We conclude in Sec. \ref{sec:conclusions}.
\section{Hamiltonian and Measurement Protocol}
\label{sec:hamiltonian_and_protocol}
    We consider an Anderson-localized chain with nearest-neighbour hopping described by the Hamiltonian
	\eq{
	\hat{H} = \sum_{i=1}^{L-1}[\ket{i}\bra{i+1}+\mathrm{h.c.}] +\sum_{i=1}^L \varepsilon_i\ket{i}\bra{i}\,,
	\label{eq:ham}
	}
	where $\ket{i}$ denotes a state perfectly localized at site $i$ and the disordered onsite potentials, $\{\varepsilon_i\}$, are randomly drawn from a gaussian distribution with zero mean and standard deviation $W/\sqrt{3}$. Each eigenfunction \(\ket{E}\) of this Hamiltonian is characterized by a center site \(j_0\) and a localization length \(\xi\), \(\bra{i}\ket{E} \propto \exp(-|i - j_0| / \xi)\) \cite{Evers-RevModPhys}. The disorder average of localization lengths at the band center attributes an average localization length \(\Bar{\xi}\) to a given disorder strength \(W\).

	We initialise the state at time $t=0$ at site $i_0$, $\ket{\psi(t=0)} = \ket{i_0}$, and study the dynamics following a discrete-time-evolution protocol. The protocol consists of two steps. In the first step, the state is subject to unitary evolution for a time interval $\Delta t$ with the Hamiltonian \eqref{eq:ham}. In the second step, with probability $p$, a \emph{single projective} measurement of the number operator is done at a randomly chosen site. Formally,
	\eq{
	\ket{\psi(t)} = \hat{\mathcal{M}}_t \ket{\psi(t^-)} = \hat{\mathcal{M}}_t e^{-\mathrm{i} H\Delta t}\ket{\psi(t-\Delta t)}\,,
	}
	where $\hat{\mathcal{M}}_t$ encodes the measurement at time $t$ and $t^-=t-0$ refers to the time just before the time of measurement $t$. The operator carries an index $t$ to make explicit the fact that it depends on time. In a given realisation of the dynamics, we denote the site that is measured at time $t$ as $i_t$ and the sequence of sites $i_{\Delta t},i_{2\Delta t},\cdots,i_{{N\Delta t}}$ as the \emph{measurement path}, where \(N\) is the total number of measurements. At any time step, the measurement site is chosen randomly and equiprobably from the entire chain.

	In the event that a measurement does happen, say at $i_t$, there are two possibilities. One is that the particle gets detected (the detector ``clicks''), in which case the state collapses into a perfectly localized (on a single site) state 
	\eq{
	\ket{\psi_{\rm c}(t)}=\ket{i_t}.\label{eq:psic}
	} 
	This happens with a probability $\rho_{i_t}(t) \equiv \vert\psi_{i_t}\vert^2$ given by the Born rule, where  $\psi_{i_t} = \langle i_t|\psi(t^-)\rangle$.
	The second possibility, with probability $1-\rho_{i_t}(t)$, is that the particle does not get detected (detector does not click -- ``no-click'' event), in which case the wavefunction develops a hole at $i_t$, such that 
	\eq{
	\ket{\psi_{\rm nc}(t)} &= \frac{1}{\sum_{i\neq i_t}|\psi_{i}(t^{-})|^2}\sum_{i\neq i_t}\psi_{i}(t^{-})\ket{i}\,\label{eq:psinc}\\
	&\propto \left[1 - \ket{i_t} \bra{i_t}\right]\ket{\psi(t^-)}.
	}
	For a given measurement path specified by the locations and times of the projective measurements, we define the sequence of clicks and no-clicks as the \emph{outcome sequence}.

	The above makes clear the various sources of stochasticity (both spatial and temporal) in the dynamics, namely, the disorder realisation, the measurement path, and the outcome sequence. For a given disorder realisation and \(i_0\), a particular measurement path \(i_t\) and a particular outcome sequence \(n_t \in [0, 1]\) together  define uniquely the \emph{quantum trajectory} \(\{t, i_t, n_t\}\). It is useful to introduce notation for averages over each of these stochasticity sources as 
	\begin{itemize}
		\item $\langle{\cdots}\rangle_\mathrm{d} \equiv$ average over disorder realisations,
		\item $\langle{\cdots}\rangle_\mathrm{p} \equiv$ average over measurement paths,
		\item $\langle{\cdots}\rangle_{\rm o} \equiv$ average over outcome sequences.
	\end{itemize}    
    These averages are defined by varying the corresponding quantities between runs of the protocol.

    In the following, we consider the disorder strengths in the interval \(W \in [2, 10]\), corresponding to average localization lengths, $\overline{\xi}$, in the range \(0.9 \lesssim \bar{\xi} \lesssim 7\). The system size is chosen to be larger than the localization length for any of the disorder strengths, \(L \gg \bar{\xi}\), to avoid finite size effects. Furthermore, we consider long unitary time evolution intervals \(\Delta t / p \gtrsim \bar{\xi}\), avoiding the quantum Zeno effect--the particle has typically enough time to spread over its localization length between two click events. 
    
   \begin{figure}[tb!]
        \centering
        \includegraphics[width=\linewidth]{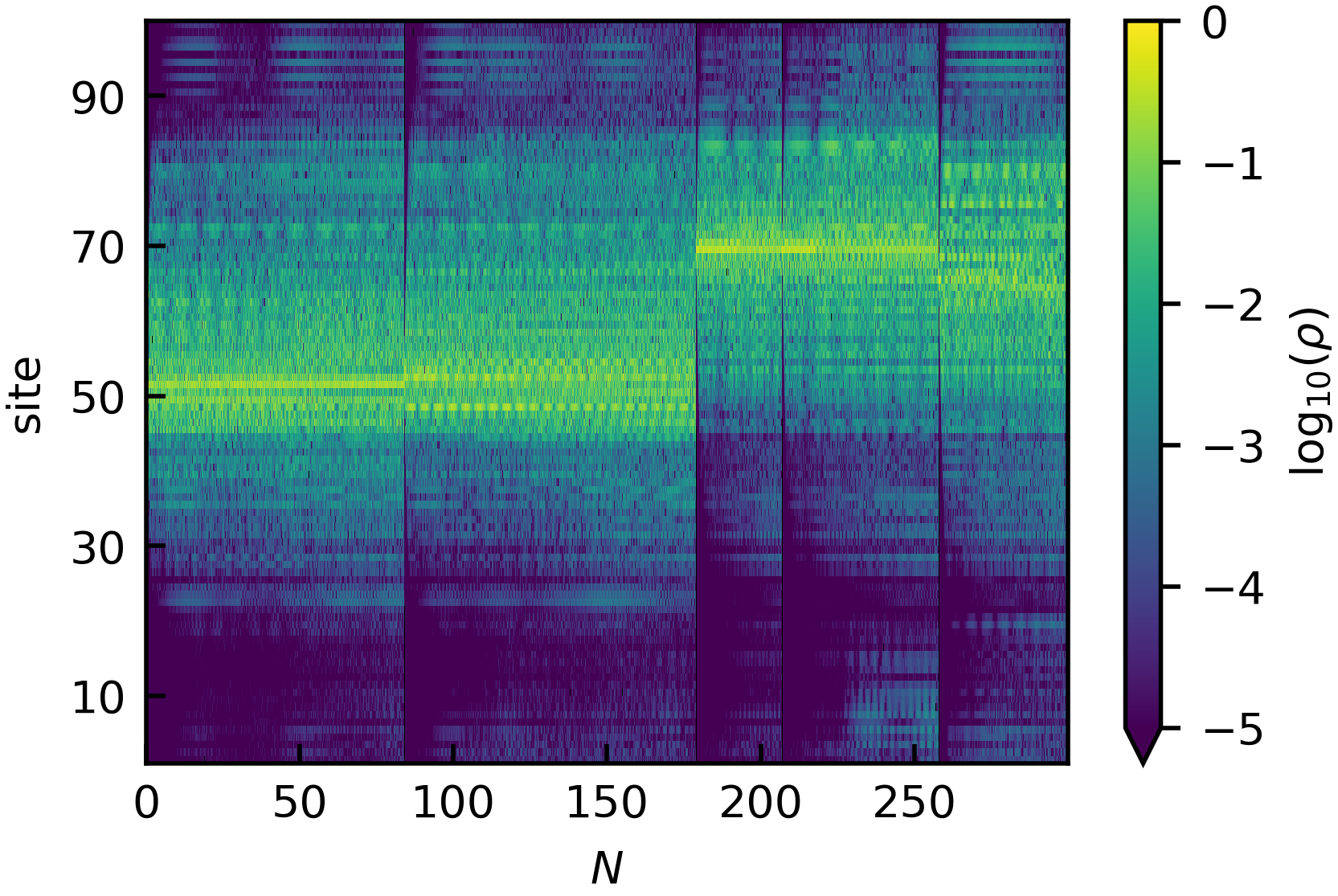}
        \caption{Time evolution of the probability density starting from the initial state \(\ket{51}\) in a system of \(100\) sites with \(\bar{\xi} \approx 7\), \(\Delta t = 10\) and \(p = 1\). We show \(100\) intermediate times for each step of the protocol. We observe both no-click and click events:
    	For  measurements at \(N = 84, 179, 207, 258\) clicks occur at sites \(52, 69, 69, 65\) and the particle is thus projected to these positions. The dark spots correspond to the no-click outcomes.}
        \label{fig:density_evolution_example}
    \end{figure}
    \section{Particle trajectories and Observables}
    \label{sec:particle_trajectories}

  To first get a broadbrush view of the phenomenology of the dynamics, we simulate a particular quantum trajectory and plot $\rho_i(t)=|\psi_i(t)|^2$ as a heatmap in the $(i,t)$ space, as shown in Fig.~\ref{fig:density_evolution_example}. The state is initialised at the centre of the chain, and we consider $W=2$, \(\Delta t = 10\) and $p=1$ as an example. The following features are of note in the dynamics: (i) the wavefunction remains reasonably localized around its localization centre throughout the dynamics; (ii) the profile of the wavefunction is quite robust to a series of several no-click events; (iii) no-click events are much more likely than click events.

   Much of these can be argued for based on the fact that the unitary dynamics between the measurements is governed by an Anderson localized Hamiltonian. This mandates that in between measurements the state can only spread within a region of size of order of $\overline{\xi}$ on average around the localization centre. Note that our wavefunction does not exactly correspond to an eigenfunction at finite disorder, because we initialize on / project to individual sites, which share overlap with \(\mathcal{O}(\bar{\xi})\) sites. For this reason, the wavefunction does not necessarily have one bright center. 
   
   If the consequent measurement is near the localization centre, it is highly likely that the detector clicks and the state gets further localized to a single site. On the other hand, if the measurement is away from the localization centre, then the probability of a click is exponentially suppressed in the distance from the localization centre. This, in turn, leads to a concomitantly small perturbation to the wavefunction in the most likely case of a no-click event and, hence, to the robustness of the state to individual no-click events. Finally, note that the probability of a click decreases with increasing system size. This can be understood as follows: the probability of a click, if the measurement is at site $i$, is simply $|\psi_i|^2$. However, since every site is equiprobable for it to be measurement site (with probability $pL^{-1}$), the probability that we have a click measurement outcome is $pL^{-1}\sum_i |\psi_i|^2 = pL^{-1}$, which is extremely small for large $L$. This also suggests that the average time interval between two click outcomes is \(L \Delta t / p\). In combination with the typical smallness of the perturbation due to one no-click, this implies the robustness of the wavefunction to typical no-click sequences. 
    
   The upshot of all of the above is that, since the wavefunction behaves like a semiclassical wavepacket at all times, its dynamics can be effectively characterised quantitatively by the moments of the wavepacket which we define as 
   \eq{
   r_q(t) = \sum_{i=1}^L i^q |\psi_i(t)|^2\,.
   \label{eq:rq}
   }
   For instance, $r_{q=1}(t)$ denotes the expectation value of the position of the wavepacket (which we refer to as \textit{particle trajectory}), $r_2(t)-r_1^2(t)$ denotes the width of the wavepacket for a given realisation and so on. 

\begin{figure}[tb!]
        \centering
        \includegraphics[width=\linewidth]{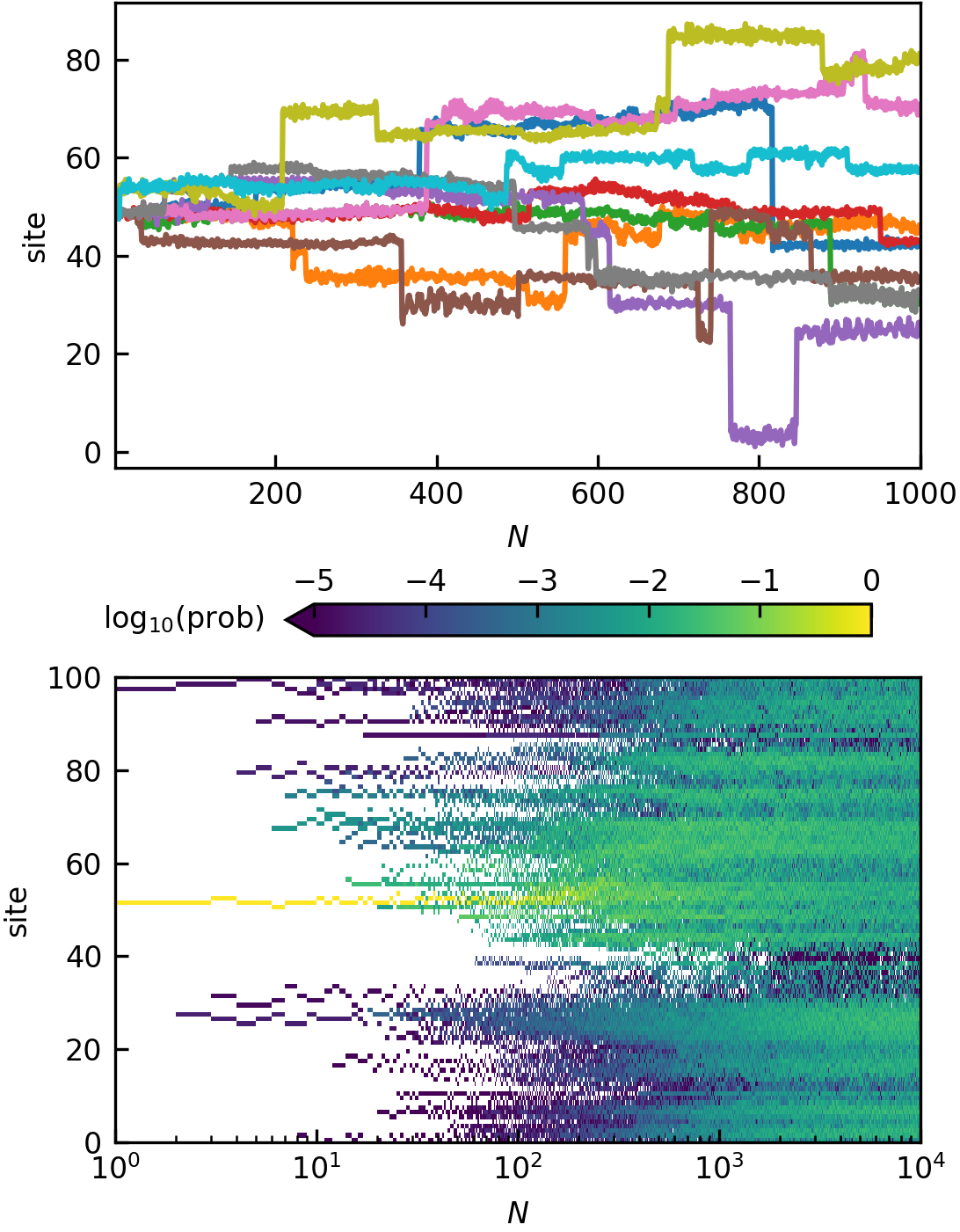}
        \caption{Upper panel: Position expectation values \(r_1(t)\) as a function of time in a system of \(L = 100\) sites, with \(\bar{\xi}(2) \approx 7\), \(p = 1\) and \(\Delta t = 100\), after initialization at \(i_0 = 51\). Differently colored lines correspond to different sequences of measurement outcomes. Disorder realization and measurement path are fixed. Lower panel: Color coded probability to have a particle trajectory at a given site after \(N\) measurements, for a particular random measurement sequence and disorder realization. Same parameters as in the upper panel. The particle trajectory positions were rounded to the nearest integer to obtain this plot.}
        \label{fig:individual_trajectories}
    \end{figure}

   Before analyzing different averages of $r_q(t)$ in the following section and motivating their physical meaning, we explain some of the features of $r_1(t)$ for a typical trajectory shown in Fig.~\ref{fig:individual_trajectories}. For all data therein we consider $W=2$, $p=1$, $\Delta t=100$, and the disorder realisation and measurement path are held fixed. The upper panel shows $r_1(t)$ for different outcome sequences (represented by different colours). Note that for large intervals of time, $\mathcal{O}(L/p)$, $r_1(t)$ only fluctuates weakly around some value -- this is a manifestation of the fact that each of the measurements are no-click events or click events very close to / at the center site of the wavefunction. Importantly, the probability of such ``zero-distance" jumps depends on the disorder strength, which thus also impacts the typical time scale between two visible jumps. The weak fluctuations are brought about by the unitary quantum evolution of the state with the hole punctured by the no-click events. However, every $\mathcal{O}(L/p)$ time instance on average, the detector clicks far from the previous center, and $r_1(t)$ jumps to the new location.

   This can be illustrated further by considering the probability distribution of the wavepacket's expected position over all outcome sequences. We introduce 
   \eq{
   \Pi_i(t) = \frac{1}{N_\mathrm{o}}\sum_{\mathrm{o}}\delta_{[r_1^{\mathrm{(o)}}(t)],i}\,,
   }
   where $[r_1^\mathrm{(o)}(t)]$ denotes the wavepacket's expected position for outcome sequence $\mathrm{o}$---rounded to the nearest integer---and $N_\mathrm{o}$ is the number of such sequences considered. The results for $\Pi_i(t)$ as a heatmap in the $(i,t)$ space are shown in the lower panel of Fig.~\ref{fig:individual_trajectories}. The key point that should be noted from the data is that the distribution of $\Pi_{i}(t)$ keeps getting broader in $i$ as time progresses. This can be attributed to the click events, albeit rare. This is most noticeable at early times in the figure, where a click event creates a new stream of finite $\Pi_{i}(t)$, which is spatially separated from other regions of finite $\Pi_{i}(t)$. A click event displaces the localization centre of the wavepacket. The probability of having a given distance between the two localization centres naturally decreases exponentially with the distance; nevertheless, averaged over outcome sequences, these rare click events can delocalize the state in an average sense.

    In order to quantify the spread of the wave packets across the system, we introduce the displacement
    \begin{align}
	    \Delta_{\mu}(t) &= \left[\langle r_2(t)\rangle_{\mu} - \langle r_1(t)\rangle_\mu^2\right]^{1/2}\,,
	    \label{eq:Delta}
    \end{align}
    where $r_q(t)$ is defined in Eq.~\eqref{eq:rq}.
    Here and in what follows $\mu$ denotes averages over runs and can be ``d'', ``o'', or ``p'', or a combination thereof. For example, \(\Delta_{\rm o, p, d}\) denotes \(\Delta\), averaged over outcomes, paths, and disorder realizations. The index \(\mu\) on the left-hand side indicates, that the quantity may depend on the type of the performed averages. The displacement \(\Delta\) captures contributions from the quantum spreading of individual wave functions due to time evolution, as well as the spreading of the ensemble of particle trajectories over the system.
    If strongly localized trajectories spread equally over the entire chain, the spread is given by \(\Delta(t) = L / \sqrt{12}\) for \(L \gg 1\).

  While $\Delta(t)$ defined in Eq.~\ref{eq:Delta} has major contributions from the classical spreading of the wavepacket due to the stochasticity in measurement paths or outcome sequences, we show in Sec.~\ref{sec:particle_trajectories} that, quantum-mechanically, the wavefunction remains reasonably localized within each trajectory, see Fig. \ref{fig:complete_outcome_average}. 
  
  To quantify this,
    we introduce as a second observable the ``effective localization length"   
     \begin{align}
        \xi^{\rm eff}_{\mu}(t) &= \sqrt{\left\langle \left\langle \left[\hat{x} - r_1(t) \right]^2 \right\rangle_{{\psi}(t)}\right\rangle}_\mu,
    \end{align}
    where the notation $\langle\ldots\rangle_{{\psi}(t)}$ denotes the quantum-mechanical average with the wavefunction $\psi$ at time $t$.
    The quantity \(\xi^{\rm eff}\) was also studied in Ref.~\cite{effective_localization_length} to find the distribution of localization lengths of eigenstates in Anderson Hamiltonians. We use it as a dynamical definition of the localization length, sensitive to changes to the shape of the wave packet due to the interplay between disorder-induced localization and measurements. For point-like densities \(\xi^{\rm eff} = 0\), for a localized wavefunction with localization length \(\xi\) it holds that \(\xi^{\rm eff} = \xi / \sqrt{2}\) (if \(L\gg \xi\) and \(\xi \gtrsim 1\)).
    Note that \(\xi^{\rm eff}(t) = \Delta(t)\), if no average over the measurement runs is performed. In general, the difference between the displacement and the effective localization length captures the classical spread of particle trajectories \footnote{This relation holds true if \(r_1\) is calculated as in a system with open boundary conditions, according to Eq.~\eqref{eq:rq}. Using periodic boundary conditions, \(r_1\) needs to be defined more carefully~\cite{com_periodic_bcs} to give correct values for wave functions close to the boundaries. In general, as long as the trajectory spread is much smaller than the system size, the boundary conditions make no difference, but if there are many trajectories close to periodic ``boundaries'', the definitions of all observables must be adapted carefully. Since the numerical results shown in Sec. \ref{sec:measurement_delocalization} were obtained with periodic boundary conditions, we exercised caution. In most cases, the shift of the wave function is much smaller than the system size such the given definitions can be employed. When we observe saturation of \(\Delta\), we calculate the position average as described in Ref. \cite{com_periodic_bcs}, and obtain \(\xi^{\rm eff}\) by first shifting the center of the wave function to the center site of the system. In this case, \(\Delta^{\rm class}\) should be explicitly calculated from \(\langle r_1^2\rangle - \langle r_1\rangle^2\)},
    \begin{align}
        \sqrt{\Delta^2_\mu(t) - [\xi^{\rm eff}_{\mu}(t)]^2} &= \sqrt{\langle r_1^2(t) \rangle_\mu - \langle r_1(t) \rangle_\mu^2}\notag \\
        &=: \Delta^{\rm class}_{\mu}(t) \in \mathbb{R}. \label{eq:classical_displacement}
    \end{align}
    We illustrate this observable in Fig. \ref{fig:all_averages}.

    During our simulations, we calculate the observables after the unitary time evolution intervals, immediately before the measurements. Therefore, one-site states resulting from click events, as well as holes from no-click events are spread out by the unitary evolution before the observables are calculated. In this convention, a click event contributes \(\mathcal{O}(\Bar{\xi})\) to the effective localization length, just like a quantum trajectory from \(p = 0\) evolution. This is desirable, since a click outcome resets the time evolution of \(\xi^{\rm eff}\), starting again from a one-site state, possibly on a different site. Calculating \(\xi^{\rm eff}\) immediately after the measurement, one would trivially get zero after a click event. Hence, \(\xi^{\rm eff}\) is calculated immediately before the measurement.

    \section{Measurement-induced delocalization}
    \label{sec:measurement_delocalization}
    In the previous section, we introduced the observables \(\Delta(t)\)---quantifying the spread of the particle trajectories---and the effective localization length \(\xi^{\rm eff}(t)\). In the absence of measurements, these observables assume their final values within \(\mathcal{O}(\bar{\xi})\) hopping times, as a consequence of  eigenfunction's localization in the Anderson chain. In this case, both observables measure the usual disorder localization length, as the particle remains confined to its initial position. In the following, we quantify the impact of uniformly distributed measurements on the dynamics of the system through these observables.
    \subsection{Different averages}
    \begin{figure}[tb!]
        \centering
        \includegraphics[width=\linewidth]{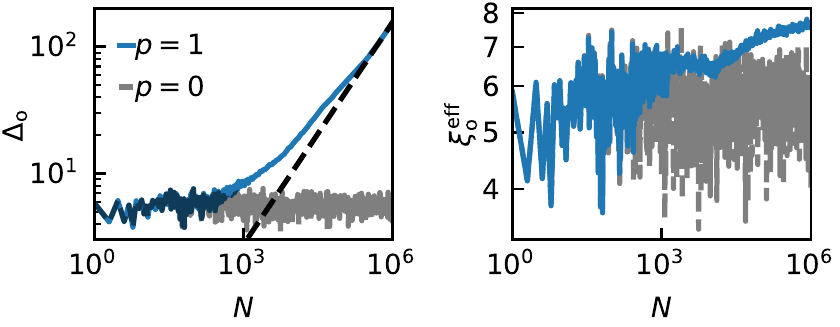}
        \includegraphics[width=\linewidth]{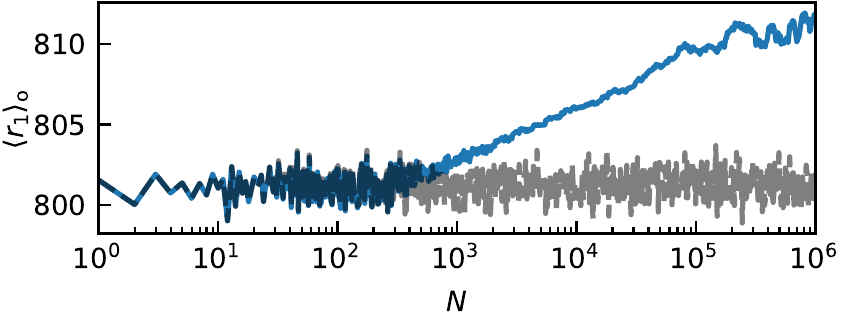}
        \caption{Upper panels: Displacement (left) and effective localization length (right) for a system of \(L=1600\) sites, with fixed disorder realization (\(\bar{\xi} \approx 7\)), \(\Delta t = 100\) and fixed random measurement path and initial site \(i_0 = 801\). Time is measured in units of \(\Delta t\); \(t = N \Delta t\). The data is averaged over \(4 \cdot 10^3\) measurement outcome sequences. The dashed line in the left panel corresponds to a power-law with \(\Delta(t) \propto t^{0.59}\). Lower panel: Position expectation value in the same system. In all panels grey (blue) lines represent \(p = 0\) (\(p = 1\)).}
        \label{fig:complete_outcome_average}
    \end{figure}
    Figure~\ref{fig:complete_outcome_average} shows the observables \(\Delta_{\rm o}\), \(\xi^{\rm eff}_{\rm o}\) and \(\langle r_1 \rangle_{\rm o}\) obtained from an outcome average over \(4 \cdot 10^3\) runs in a system of \(L  = 1600\) sites, with a fixed disorder realization for \(W = 2\) corresponding to \(\bar{\xi} \approx 7\) and fixed random measurement path with \(p = 1\) (blue lines) and \(\Delta t = 100\). Grey, dashed lines show the same observables in the non-measured case \(p=0\) for comparison. Throughout section \ref{sec:measurement_delocalization}, we use periodic boundary conditions.
    
    Let us first consider the outcome-averaged position expectation value: At \(p=0\) this value fluctuates by \(\mathcal{O}(\bar{\xi})\) sites around the initial position. These fluctuations are due to the unitary time evolution, mediating between \(\ket{i_0}\) and \(\mathcal{O}(\bar{ \xi})\) neighboring sites through the localized eigenfunctions. In contrast, at \(p=1\), the fluctuations are less pronounced, as they average out over different quantum trajectories. At \(N \sim L = 1600\), however, the average position starts to slowly drift away from the initial position, reaching \(i \approx 810\) at \(N \approx 10^5\). The drift velocity is very small \(v \approx 1 / 10^4 \ll 1\). At \(N \geq 10^5\) the average drift continues even more slowly, while fluctuations of magnitude \(\mathcal{O}(1)\) emerge.

    Since \(p=0\) corresponds to a single quantum trajectory, only the spread of the wavefunction contributes to the displacement \(\Delta\). Accordingly, \(\xi^{\rm eff}\) and \(\Delta\) are equal in this case, and \(\Delta = \xi^{\rm eff}\sim \bar{\xi}\).
    This is similar for the measured case up to \(N \ll L / p = 1600\), where again both quantities behave similarly (not shown), since all quantum trajectories away from the initial position are very unlikely---few click events occur up to this point and there are also few impactful no-click events up to \(N \sim L / (p\xi) \approx 250\). In consequence, there is no classical contribution to \(\Delta_{\rm o}\) from the displacement of the wave packet center, and the shape of the  wave packet is largely determined by the unitary time evolution. At \(N \sim L\), however, \(\xi^{\rm eff}_{\rm o}\) at \(p = 1\) is increased compared to non-measured \(\xi^{\rm eff}\), while \(\Delta_{\rm o}\) continues to grow as an approximate power law \(\propto t^{\gamma}\) with \(\gamma \approx 0.59\) (close to the diffusive exponent \(\gamma_{\rm diff} = 1 / 2\)). 
    
    At \(N = L\) for \(p = 1\), we expect \(\mathcal{O}(1)\) click-events for each trajectory, shifting the positions of the wave packets; and no-click events close to the center of the wave packets, redistributing the weight which is projected away from the measured site. Importantly, however, \(\xi^{\rm eff}_{\rm o} \ll L\), validating the picture of an effective localization length that is still defined in the presence of measurements.
    Considering the entire time interval, we notice slow fluctuations in the local power law exponent of \(\Delta_{\rm o}\); \(\xi^{\rm eff}_{\rm o}\) reaches a local minimum at \(t \sim 10^4\) and slightly increases at later times.

    If we redraw the measurement path between runs, in addition to considering random outcome sequences for every run,
    the features of the plots are largely similar to the exclusively outcome averaged case (not shown). This is expected even for \(N \gg L\), when measurements play an important role. The outcome-average alone spatially separates the particle trajectories over time, and at this point different trajectories experience independent measurement locations in their vicinities anyway. There may be subtle differences at intermediate time scales, where many trajectories are still overlapping. However, we are mostly interested in the long-time behaviour.
    \begin{figure}[tb!]
        \centering
        \includegraphics[width=\linewidth]{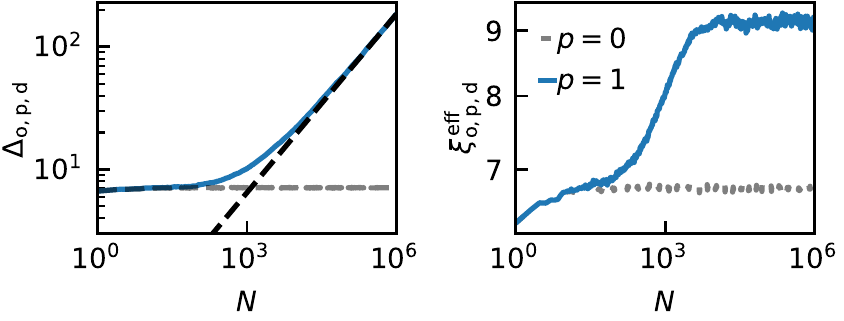}
        \includegraphics[width=\linewidth]{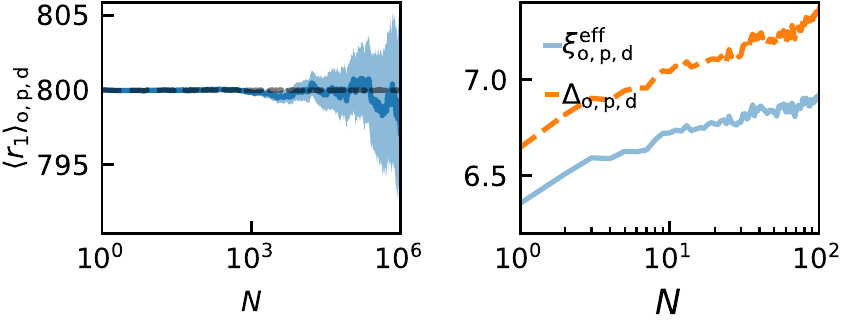}
        \includegraphics[width=\linewidth]{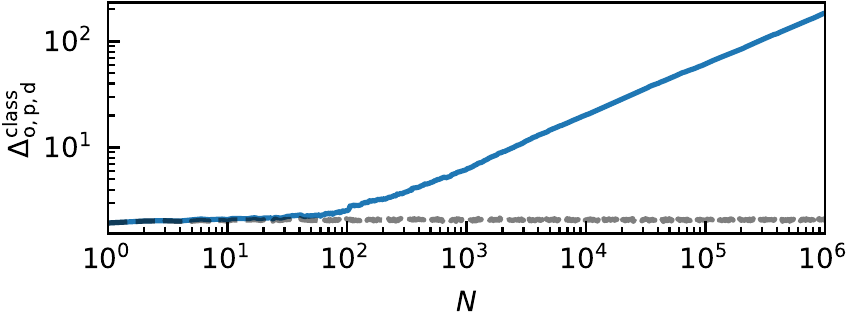}
        \caption{Upper panel: Displacement (left) and effective localization length (right) for a system of \(L=1600\) sites, with \(\bar{ \xi} \approx 7\) and \(\Delta t = 100\). The data is averaged over \(4 \cdot 10^3\) runs, with random measurement outcomes, paths, and disorder realizations and initial site \(i_0 = 801\). The black dashed line in the left panel corresponds to a power-law with \(\Delta(t) \propto t^{0.49}\). Blue and grey lines correspond to \(p =1\) and \(p = 0\) respectively in all panels. Middle panel: Position expectation value in the same system with two-sigma finite sample size error estimation from a bootstrap procedure (shaded region) (left). Comparison between displacement and effective localization length during the initial evolution (right). Lower panel: Classical displacement \eqref{eq:classical_displacement}.}
        \label{fig:all_averages}
    \end{figure}

    Finally, we consider the observables \(\Delta_{\rm o, p, d}\) \(\xi^{\rm eff}_{\rm o, p, d}\), \(\langle r_1 \rangle_{\rm o, p, d}\) and \(\Delta^{\rm class}_{\rm o, p, d}\) with averages over outcomes, paths and disorder at the same parameters \(W = 2\), \(L = 1600\), \(p = 1\), \(\Delta t = 100\) (Fig. \ref{fig:all_averages}). In addition, \(\Delta_{\rm o, p, d}\) and \(\xi^{\rm eff}_{\rm o, p, d}\) are plotted together in the lower right panel, to demonstrate the difference between \( \Delta^{\rm class}_{\rm o, p, d}(t)\) \eqref{eq:classical_displacement} and \(\Delta_{\rm o, p, d}(t)\). Averaging over disorder realizations and paths, we remove all spatial inhomogeneities from the averaged quantities. As a result, \(\langle r_1 \rangle_{\rm o, p, d}\) is constant up to finite sample fluctuations, also in the measured case. The remaining fluctuations can be explained as follows by the finite sample average: \(\Delta^{\rm class}_{\rm o, p, d}(t)\) growing with time means that the mean position \(r_1\) of a given trajectory can be considered a random number sampled from an increasingly broad distribution (see Sec. \ref{sec:random_walk} for details). Thus, we estimate the magnitude of the finite-sample fluctuations as a function of time by calculating the variance \(\sigma^2\) in a large sample of averages of \(4 \cdot 10^3\) random numbers per sample from a gaussian distribution of standard deviation \(\Delta^{\rm class}_{\rm o, p, d}(t)\). We observe, that \(\langle r_1(t)\rangle_{\rm o, p, d}\) is always within \(\pm 2 \sigma\) (blue shaded area) from zero.
    
    The obtained exponent for the spread \(\gamma = 0.49\) is very close to the diffusive value. Contrary to the previous cases, \(\xi^{\rm eff}_{\rm o, p, d}\) saturates after \(\mathcal{O}(10^3)\) measurements to a value that does not further change with time. This is because the introduction of an average over disorder realizations removes, right from the outset, correlations between different runs of the simulation. Regarding the classical displacement, we note that \(\Delta^{\rm class}_{\rm o, p, d} \approx 1 > 0\) for \(p = 0\) as well as \(p= 1\) (with \(N \lesssim 10^2\)). This is due to the spread of the initial wave function across \(\mathcal{O}(1)\) sites around \(i_0 = 801\). In different disorder realizations, the center position \(r_1\) slightly varies, reflecting in a finite value of \(\Delta^{\rm class}_{\rm o, p, d}\). At later times, the \(p = 0\) curve remains at this initial value, while the \(p = 1\) curve grows according to the measurement induced spread of the particle trajectories. Fitting \(\Delta^{\rm class}_{\rm o, p, d}\) at \(N \geq 5 \cdot 10^5\) to a power law, we obtain again \(\gamma = 0.49\). Importantly, the classical contribution to \(\Delta_{\rm o, p, d}\) determines its behaviour in the long time limit, since the effective localization length converges to a system size independent value, while the trajectories spread over the entire system.

    The average number of measurements required to displace the wave packet center from a specific site is determined by the localization lengths of the eigenfunctions peaked close to this site, which are, in turn, determined by the disorder realization.
    If disorder is not averaged over, a region of small localization lengths (``traps") slows down all trajectories passing through that region. Such traps can lead to drifting of \(\langle r_1 \rangle_{\rm o, p}(t)\) (partially blocking transport on one side of the system), as well as to fluctuations in \(\Delta_{\rm o, p}(t)\) (traps slow down the average spread) and \(\xi^{\rm eff}_{\rm o, p}(t)\) (the trap corresponds to small \(\xi\) and, thus, small \(\xi^{\rm eff}\)). On the other hand, regions of large \(\xi\) can speed up the spread and lead to upwards fluctuations in \(\xi^{\rm eff}_{\rm o, p}\).
    In the thermodynamic limit \(L, N \rightarrow \infty\), we expect these effects to vanish, even if the observables are only averaged over outcomes, as the trajectories become increasingly spatially distributed and thus less correlated. This suggests, that the diffusive power law exponent obtained by performing all averages should also be seen at long times, if only an average over outcomes is performed. However, our numerics do not probe the corresponding time scales.  
    Upon disorder averaging, there is an immediate average over different \(\xi\) at every point in time, since the observables are averaged with wave functions from different disorder realizations, leading to \(\gamma \approx 1 / 2\) for a sufficiently large sample average. In contrast, if no average over disorder realizations is performed, it takes much longer to reach the diffusive limit. Due to traps, deviating exponents are observed for different disorder realizations on intermediate time scales. 
    
    Based on the approximation that the trajectories spread diffusively due to click-events, we can estimate the number of measurements required to achieve a sufficient effective average over different localization lengths without performing the disorder average. From Fig. \ref{fig:all_averages}, we conclude that \(\sim 1000\) independent sites are sufficient for the effect of traps to be averaged out. Therefore, having \(\Delta_{\rm o, p} \sim 10^3 \bar{\xi}\) without disorder averaging should facilitate a similar average. This corresponds to \(t' \sim 10^{6}\), where \(t' = N p / L\) is the expected number of click events. We find \(N \sim 10^6 L / p\), where we have to set \(L \gtrsim 10^3 \bar{\xi}\) to avoid finite-size effects. We find \(N \sim 10^9 \bar{\xi} / p\) with an additional numerical \(\bar{\xi}\)-dependent factor taking into account zero-distance jumps. From this estimate, it is clear that we would have to go to much larger numbers of measurements to find diffusive behaviour without averaging over disorder realizations. 
    
\begin{figure}[tb!]
        \centering
        \includegraphics[width=\linewidth]{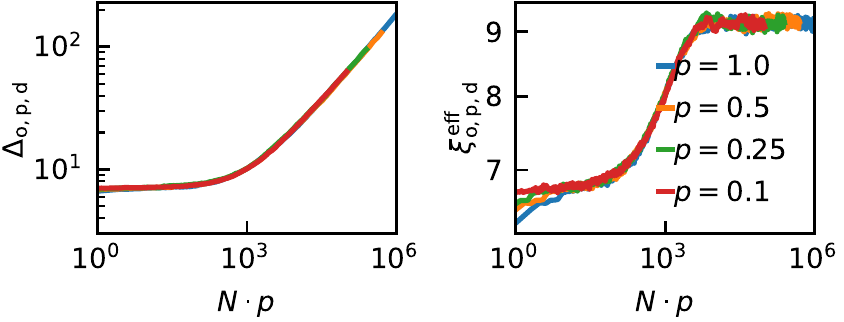}
        \caption{Displacement (left) and effective localization length (right) for a system of \(L=1600\) sites, at \(\bar{ \xi} \approx 7\) and \(\Delta t = 100\) for different measurement probabilities. Each curve is averaged over \(4 \cdot 10^3\) runs, with random measurement outcomes, paths, and disorder realizations. The data is plotted over the expected number of measurements \(N \cdot p\).}
        \label{fig:different_measurement_probabilities}
    \end{figure}

    In Fig. \ref{fig:different_measurement_probabilities}, we compare \(\Delta_{\rm o, p, d}\) and \(\xi^{\rm eff}_{\rm o, p, d}\) at different measurement probabilities in order to demonstrate that the measurement frequency only rescales the time axis and has no impact on the diffusive exponent, as long as \(\Delta t \gg \xi\). For this purpose, we plot each quantity as a function of the expected number of measurements 
    \(N \cdot p\) on top of each other. Indeed, after an initial phase corresponding to few measurements, the curves lie on top of each other.
    
    \subsection{Length and time scales}
    
    \begin{figure}[tb!]
        \centering
        \includegraphics[width=\linewidth]{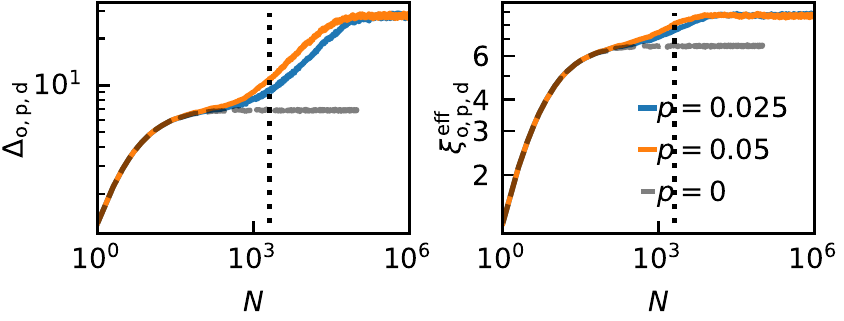}
        \caption{Displacement (left) and effective localization length (right) in a system with \(\bar{ \xi} \approx 7\) and \(L = 100\). The data is averaged over \(4 \cdot 10^3\) runs, with random outcomes, disorder realizations and measurement paths (\(\Delta t = 1\) here). The vertical dotted lines mark one expected click event, \(N_{1\rm c} = L / p\) for \(p = 0.05\).}
        \label{fig:two_plateaus}
    \end{figure}

    In order to separate the influence of measurements on the average quantities from the non-measured time evolution, we consider small measurement probabilities \(p \in \{0.025, 0.05\}\) in a system with \(L=100\) and \(\bar{ \xi}\approx 7\) (\(p=0\) is shown for reference), averaging over \(\mathcal{O}(10^3)\) runs, see Fig.~\ref{fig:two_plateaus}. We choose \(\Delta t  =1\), which does not come with the Zeno effect, since \(\Delta t / p \gg \bar{\xi}\). In order to avoid the related intermediate-scale effects, we average not only over outcomes and measurement paths, but also over disorder realizations.
    Because of the small measurement probability, it takes many steps of the time evolution protocol for the measurements to show a pronounced effect on the system, leading to a separation of the initial time scale, where the observables basically behave as in the absence of measurements (gray, dashed lines), from the time scale, where the effect of measurements sets in. 
    Specifically, for these parameters of the protocol, \(N = L / (\bar{\xi } p) \sim 300\) steps are required until one measurement within the localization radius has taken place on average, and about \(N = L / p \sim 2000\) steps (black dotted lines) until one click is encountered. Consequently, a difference between the \(p = 0\) and \(p = 1\) curves becomes apparent between these time steps.
    
    As we observed before, \(\xi^{\rm eff}_{\rm o, p, d}(t)\) increases to its saturated value \(\bar{ \xi} < \xi^{\rm eff}(t \rightarrow \infty) \ll L\) while \(\Delta_{\rm o, p, d}(t)\) grows according to a power law. 
    The spread of the trajectories is limited by the system size, which shows as a second plateau with \(\Delta_{\rm o, p, d} \approx L / \sqrt{12}\approx 30\)---corresponding to uniform spreading over the system. Assuming diffusive spreading of trajectories, driven by click events, the corresponding scale can be estimated as \({t'}^{1 / 2} \sim L / \bar{\xi}\), where the time is counted in expected clicks \(t' = N p / L\), giving \(N \sim L^3 / (p \bar{\xi})^2 \sim 6 \cdot 10^{5}\), in agreement with the actual time of saturation to the second plateau. Since \(\xi^{\rm eff}(t \rightarrow \infty)\) is related to no-click events, the corresponding plateau sets in when there is averaging over contributions of all relevant numbers of successive no-click events. As a rough estimate, the probability of a sequence of exclusively no-click events decreases exponentially with the length of this sequence. We thus estimate \(N \sim L / (p \bar{\xi}) \sim 300\) with the numerical factor depending on the details of effective localization length's origin.
    
    In the upper panel of Fig. \ref{fig:time_and_size_scaling}, we demonstrate the scaling of \(\Delta(t \rightarrow \infty)\) (left) and \(\xi^{\rm eff}(t \rightarrow \infty)\) (right) with the size of the system for parameters \(p = 1\), \(W = 2\), \(\Delta t = 100\). In agreement with uniform spread over the entire system, we find \(\sqrt{12} \Delta(t \rightarrow \infty)(L) \approx L\) (lower left, dashed line). On the contrary, \(\xi^{\rm eff}(t \rightarrow \infty, L)\) (lower right) saturates to \(\sqrt{2}\xi^{\rm eff}(t \rightarrow \infty, L \rightarrow \infty) \approx 13\) for \(L \gtrsim 250\). This further validates that there are still localized wave functions in the presence of measurements, despite the spread of trajectories. In the lower panels, we plot the number of measurements required to reach the plateaus in \(\Delta_{\rm o, p, d}\) (left) and \(\xi^{\rm eff}_{\rm o, p, d}\) (right); confirming the cubic (linear) dependence on the system size.
    
    \subsection{Effective localization lengths}

    \begin{figure}[tb!]
        \centering
        \includegraphics[width=\linewidth]{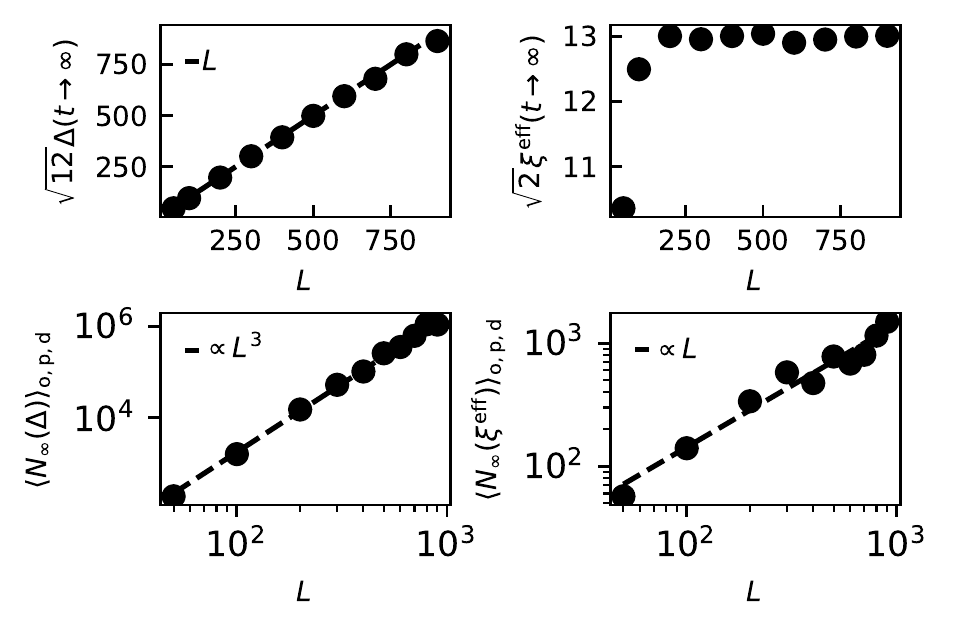}
        \caption{Upper panels: System size scaling of the time converged values \(\sqrt{12}\Delta(t \rightarrow \infty)\) (left) and \(\sqrt{2}\xi^{\rm eff}(t \rightarrow \infty)\) (right) at \(\bar{ \xi} \approx 7\), \(p = 1\) and \(\Delta t = 100\). Lower panels: Required number of measurements to reach 90\% of the plateau value for \(\Delta_{\rm o, p, d}\) (left) and \(\xi^{\rm eff}_{\rm o, p, d}\) (right) in the same system.}
        \label{fig:time_and_size_scaling}
    \end{figure}

    \begin{figure}[tb!]
        \centering
        \includegraphics[width=\linewidth]{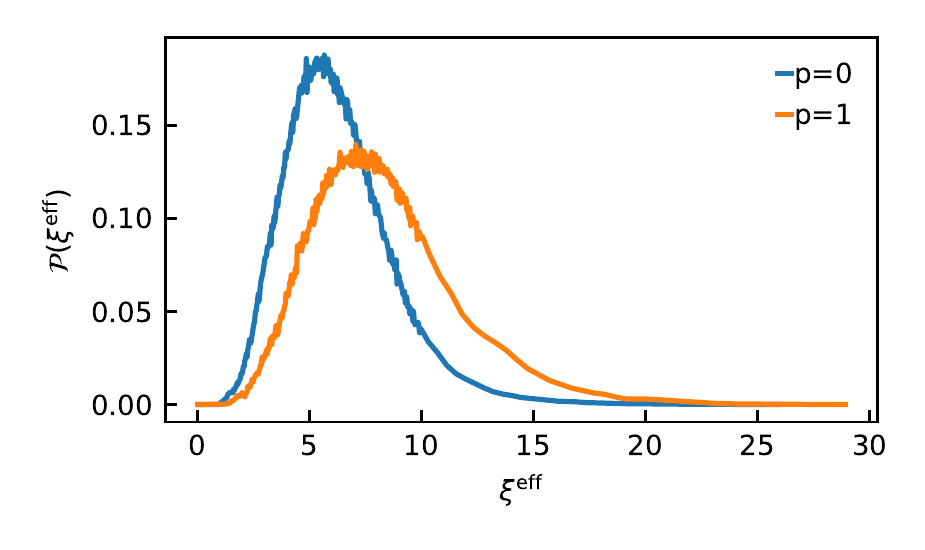}
        \includegraphics[width=\linewidth]{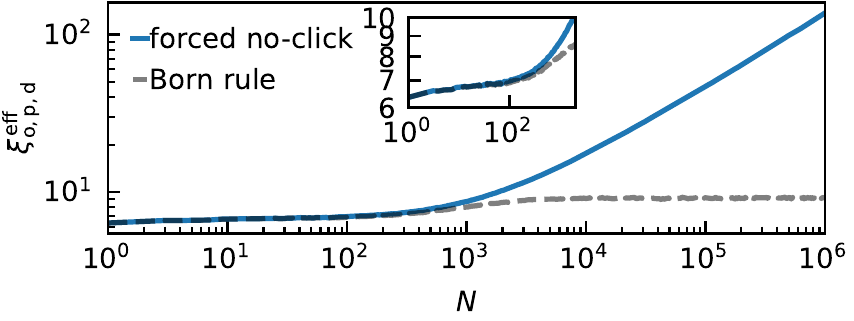}
        \caption{Upper panel: Probability distribution of effective localization lengths in measured (\(p=1\)) and non measured (\(p=0\)) system for \(\bar{\xi} \approx 7\), \(L = 300\), and \(\Delta t = 100\). These effective localization lengths were obtained in the time window \([2900, 3000]\) (where the effective localization length is converged for these parameters, see Fig. \ref{fig:time_and_size_scaling}) and averaged over \(4000\) runs with different disorder realizations and measurement paths. Lower panel: Comparison of \(\xi^{\rm eff}_{\rm o, p, d}\) with forced no-click only outcomes to usual time evolution according to our protocol, both at parameters \(L = 1600\), \(W = 2\), \(\Delta t = 100\). The inset shows a close-up of the time interval \([1, 2 \cdot 10^3]\).}
        \label{fig:xi_eff_stats}
    \end{figure}
    
    The value of \(\xi^{\rm eff}(t \rightarrow \infty)\) depends non-trivially on \(W\), \(p\), \(\Delta t\) and \(L\). In particular, \(W\) determines \(\bar{ \xi}\) and, thus, \(\xi^{\rm eff}(t \rightarrow \infty)(p = 0)\). We already observed that finite \(p\) leads to an increase in \(\xi^{\rm eff}(t \rightarrow \infty)\) in the limit \(\Delta t / p \gg \bar{ \xi }\). We do not consider the Zeno limit \(\Delta t / p \rightarrow 0\), where \(\xi^{\rm eff}(t \rightarrow \infty) \rightarrow 0\); however this limit implies that \(\xi^{\rm eff}(t \rightarrow \infty)(\Delta t / p)\) has a maximum at finite \(\Delta t / p\). In the upper right panel of Fig. \ref{fig:time_and_size_scaling}, we see the effect of \(L\) acting as an upper cutoff on \(\xi^{\rm eff}(t \rightarrow \infty)\).
    
    Concerning the effect of measurements on the effective localization length, we know that every click outcome resets \(\xi^{\rm eff} \rightarrow \xi\). A no-click outcome outside of the effective localization length has an exponentially small effect. A no-click event within the effective localization length can lead to enhancement of the wave function tails and, thus, to a slight growth in \(\xi^{\rm eff}\), resulting in \(\xi^{\rm eff}(t \rightarrow \infty) > \bar{ \xi}\).  To further illustrate the effect of measurements on the localization length, we compare in the upper panel of Fig.~\ref{fig:xi_eff_stats} the distributions of effective localization lengths in the measured and non-measured case for \(\bar{\xi} \approx 7\), \(\Delta t = 100\), \(L = 300\). In this figure, we histogrammize effective localization lengths obtained from \(4 \cdot 10^3\) random disorder realizations time evolved with our protocol, at \(p = 0\) and \(p = 1\) respectively. For every instance, we calculate \(\xi^{\rm eff}\) at 100 successive time steps, between \(2.9 \cdot 10^3\) and \(3 \cdot 10^3\). 
    
    The \(p = 0\) distribution captures the stochasticity of localization lengths in the non-measured system between different disorder realizations, which was investigated for bulk eigenfunctions in Ref.~\cite{effective_localization_length}. As expected, the measurements enhance the distribution towards larger localization lengths. Importantly however, the overall shape of the distribution is qualitatively preserved. While the distribution of inverse localization lengths of eigenstates at fixed energy is known to be gaussian \cite{Evers-RevModPhys} this only approximately describes the distributions shown in Fig. \ref{fig:xi_eff_stats} since our wave functions are given by linear combinations of a few eigenstates, and states of all energies are taken into account. Prominent features are a maximum (typical \(\xi^{\rm eff}\)) between quick decay towards small localization lengths and a long tail towards large localization lengths.
    Since in the fully averaged case the effective localization length is still larger than in the free system, we conclude that the increase of average \(\xi^{\rm eff}\) in the presence of measurements has to be connected to no-click events, since clicks reset the contribution of a wave function to the non-measured value. 
    
    To show this more explicitly, consider the lower panel of Fig.~\ref{fig:xi_eff_stats}: The grey dashed line shows the already established result for \(\xi^{\rm eff}_{\rm o, p, d}\) at parameters \(L = 1600\), \(\Delta t = 100\), \(W = 2\), which converges to \(\xi^{\rm eff}(t \rightarrow\infty) \approx 9\). The blue solid line is obtained by forcing a no-click outcome \eqref{eq:psinc} at every measured site. The inset shows a close-up of the time interval \([1, 2 \cdot 10^3]\) to demonstrate that the curves coincide in the initial phase, where the expected number of click events is low. In the forced no-click case, in contrast to the Born rule simulation, \(\xi^{\rm eff}_{\rm o, p, d}\) does not show a plateau, but continues to increase throughout the observed time window. This demonstrates again the delocalizing effect of no-click events. Without occasional click events, the wave functions would eventually completely delocalize and spread across the entire system. Note that this delocalization process could take much longer if we simply post-selected trajectories with no-clicks only. In such a trajectory the measurements would typically take place in the tail of the wave function, where the impact on the wave function is smaller. Forcing no-click outcomes, the measurement position is equally distributed along the chain. This procedure bears a certain similarity to ``forced measurements'' discussed in Ref.~\cite{forced_measurement}, but in that paper all quantum trajectories (also involving click outcomes) were forced to be equally likely. In our case, we have ``forced measurements with no-click postselection." When the Born rule is employed, delocalization induced by no-click outcomes is stopped by a single click event; as a result, the true effective localization length saturates.

    \par
    {\hphantom{a}}
    
    To summarize the above, we have found that randomly distributed measurements lead to delocalization of particle trajectories. Each projective measurement transfers the particle to the measured site. Following this process for a sufficient time span, the probability to find a particle at a particular site is equal for all sites. At long times, the spread of particle trajectories due to this process is described by a diffusive power law \(\Delta(t) \propto t^{1 / 2}\) independent of the performed averages. During this process, the wave functions are still well described by an effective localization length, as opposed to spreading over the size of the system, despite the delocalizing impact of no-click events.

    \section{Steering with measurements}
    
    \label{sec:steering}
    In the preceding section, we concluded that random measurements all over the system lead to delocalization of quantum trajectories, while almost all individual trajectories correspond to localized particles. This raises the question, if the well-defined location of the particle can be efficiently manipulated, inducing controlled transport in a ``localized" system by performing measurements according to an appropriate steering protocol.
    
    The concept of having localized trajectories with the location governed by click events can be applied to steer the particle from its initial site to a specified target, with the goal of having a click at the target. The average number of measurements required to achieve this goal defines efficiency of the measurement protocol, which dictates the measurement path. In a localized system, where only the target site is measured, the expected number of measurements would increase exponentially with the system size. Contrarily, if the wavefunction is completely delocalized, the expected number of measurements would scale linearly with system size, when measuring again only the target site. In our localized system, where all sites may be measured, we expect to find efficient (sub-exponential), non-trivial measuring strategies, since, on the one hand, quantum trajectories seem to spread over the system on a non-exponential time scale, while, on the other hand, still corresponding to localized wavefunctions.
    
    In order to investigate spatial steering, we consider the following setup. In a system of \(L\) sites we initialize the particle at site \(i_0 = 1\) and specify a target site \(i_{\rm target} = L\) (using open boundary conditions). Again, we consider large times between two measurements \(\Delta t \gtrsim \bar{\xi}(W)\), in order to avoid confining or repelling the particle through the Zeno effect. The goal is to design a measurement protocol that leads to a click outcome at the target site after as few measurements as possible. The time of arrival in a quantum lattice has also been studied in Refs. \cite{quantum_time_of_arrival, arrival_time_resetting}. Hereby, we differentiate between two types of protocols: \textit{Adaptive protocols}, which may use the readouts from every performed measurement in determining the position of the successive measurements; and \textit{blind protocols}, which may only use the readout at the target site.

    The simplest protocol one may think of is to just repeatedly measure the target site, until the particle is detected. Since this protocol requires readout only at the target site, it is a \textit{blind} protocol with the termination policy employed. Based on the localized nature of the system, this protocol typically terminates after an exponentially large number of measurements \( N \in \mathcal{O}(\exp(L / \bar{\xi}))\), upon averaging over disorder realizations. As this protocol scales exponentially with the system size, it becomes impractical (also for numerical simulations), if \(L\) is of the order of a couple of localization lengths.

    At the same time, there is a simple adaptive protocol, which is optimal in the sense that \(\langle N\rangle_{\rm o, d} \propto L\). This protocol works as follows:
    \begin{enumerate}
        \item Place the detector at \(i_1 = i_0 + 1\).
        \item Measure this site, until the particle is detected.
        \item Shift the detector by one site towards the target.
        \item Repeat steps 2 - 4.
    \end{enumerate}

\begin{figure}[tb!]
        \centering
        \includegraphics[width=\linewidth]{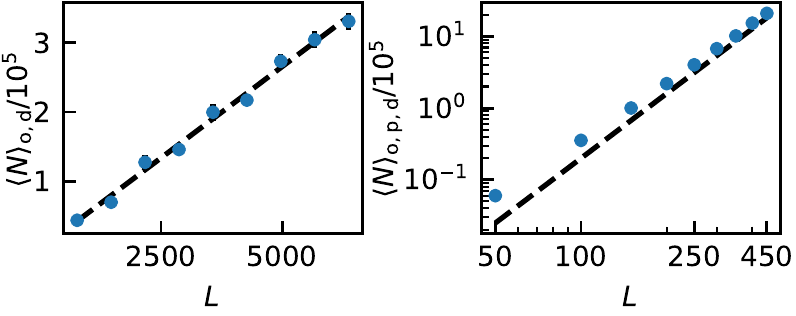}
        \caption{Scaling of adaptive readout protocol (left panel) and random blind protocol (right panel) as a function of system size, for \(W = 2\), \(\Delta t = 100\). The dashed lines show a linear fit (left panel) and the function \(f(L) \sim L^3\) (right panel) as a guide for the eye.}
        \label{fig:steering_scaling}
    \end{figure}

    Given that the expected number of measurements until the next click is finite, the total number of measurements scales linearly with the system size.
   For numerical simulations, we use a slightly improved version of this protocol: Instead of always measuring at a distance 1 from the site where the last click event took place, we randomly measure sites within one localization length of this site in the direction of the target site. This has two benefits: Local dips of the wave function as well as effects of preceding no-click events are avoided.
   Indeed, the left panel of Fig.~\ref{fig:steering_scaling} demonstrates that the simulated expected number of measurements for this protocol (blue dots) scales approximately linearly with the system size (black fit line). In this sense, ballistic transport is realized by this protocol.

   Evidently, efficient steering is possible, if the readout is always accessible. In an experiment, however, this may not be the case. Therefore, we try to find a blind protocol with an expected number of measurements that behaves polynomially with the system size.
   As an attempt to improve the runtime of the blind protocol, we perform blind measurements along the chain at random locations until the particle is detected at the target site. As we already observed, this leads to approximately diffusive spread of the trajectories and is thus much more efficient than only measuring the target site. In simulations, we are able to steer the particle to a target site at a distance of several hundred localization lengths, see the right panel of Fig.~\ref{fig:steering_scaling}. The simulation values for \(\langle N\rangle_{\rm d, p, o}\) (blue dots) scale approximately with \(f(L) \sim L^3\) (black dashed line), as expected in a diffusive system.

We have thus demonstrated a possibility of efficiently manipulating (dragging) a particle subject to a random potential in a one-dimensional chain by means of measurement-induced steering, using both passive (blind) and active (adaptive) protocols.
This type of steering can be further generalized to more sophisticated scenarios, as compared to simply moving the particle through the chain from one end to the other. In particular, one may envision manipulating several particles in the disordered background (not necessarily in a one-dimensional system) to exchange their positions and braid them by measurements.

\section{Relation to a classical random walk}
     \label{sec:random_walk}
     
    In this section, we describe how the dynamics of the spread of particle trajectories in our measured system is related to a classical random walk model. The random walk picture is useful for several reasons. For a wide variety of random walks, asymptotic properties are known, allowing us to explain the long-time behaviour of our system. The random walk language offers a simplified description of the dynamical features of the ensemble of particle trajectories, which are much more difficult to calculate analytically when taking its full quantum nature into account. Our main question is about the asymptotic behaviour of the spread \(\Delta_{\mu}(t)\) (or, equivalently, the asymptotic behaviour of \(\Delta_\mu^{\rm class}(t)\)). In Sec. \ref{sec:measurement_delocalization} we argued, that the particle trajectories spread diffusively in the long time limit, and in the following we use the random walk picture to back up this statement analytically.\par

    On the level of particle trajectories, the measurement-induced dynamics bears immediate similarity to a classical random walk. Consider a set of states \(\{i\}\) with \(i \in [1, L]\), representing the sites of the system. Approximating the position of a trajectory by the nearest site and limiting our consideration to the discrete set of time points immediately after a measurement, every particle trajectory is described by transitions \(i \rightarrow j\) with \(i, j \in [1, L]\). A natural approach is to describe the ensemble of different particle trajectories in terms of transition matrices \(M(n)\) with \(n \in [1, N]\), acting on a state \(\varrho\) with \(\varrho_i(n)\) corresponding to the probability to find a particle on site \(i\) at time step \(n\), and
    \begin{align}
        \varrho_i(n + 1) &= \sum_{j = 1}^{L}
        M_{i, j}(n) \varrho_{j}(n). \label{eq:random_walk_time_evolution}
    \end{align}
    We assume that the transitions are mediated by click events, with the transition probabilities determined by the wave functions immediately before the measurement. Indeed, we showed in Sec. \ref{sec:measurement_delocalization} that the effect of no-click events can essentially be viewed as a correction to the localization length. Since on average every \(L\)-th measurement produces a click, one time step in Eq. \eqref{eq:random_walk_time_evolution} thus implies \(L\) steps of the measurement protocol.
    
    At this point, we need to specify properties of the \(M_{i, j}(n)\), incorporating localization into the classical picture through the statistics of these matrix elements. 
    Localization implies that the transition probability decreases exponentially with the distance. Therefore, we keep only transitions over distances one and zero (distance one representing unit distance \(\mathcal{O}(\bar{\xi})\) jumps) and consider a symmetrical nearest-neighbor transition matrix
    \begin{align}
        M_{i, j}(n) &= \delta_{i, j} p_i(n) + \delta_{(i + 1), j} \frac{\bar{p}_{j}(n)}{2} + \delta_{(i - 1), j} \frac{\bar{p}_{j}(n)}{2};\\
        \bar{p}_i &= 1 - p_i \qquad i, j \in [1, L], 
        \label{eq:transition_matrix}
    \end{align}
    where respective boundary conditions should be taken into account.
    
    In this model, at every time step \(n\), the particle either remains on a given site \(i\) with \textit{waiting probability} \(p_i(n)\), or jumps with equal probabilities \(\bar{p}_i(n) / 2\) to one of the two adjacent sites. The waiting probabilities are assumed to be time independent, \(p_i(n) =: p_i \quad \forall n\) (we comment later on this assumption). They are drawn from a probability distribution \(\mathcal{P}_p(p)\) which is related to localization and determined in the following.\par
    The key insight is, that the asymptotic behaviour of the spread \(\Delta_\mu(t)\) crucially depends on the probability of zero distance jumps, corresponding to successive click events at the same site (\textit{waiting} on that site). If zero distance jumps never occurred (\(p_i = 0 \quad \forall i\)), the corresponding random walk would necessarily give diffusion.\par
\begin{figure}[tb!]
        \centering
        \includegraphics[width=\linewidth]{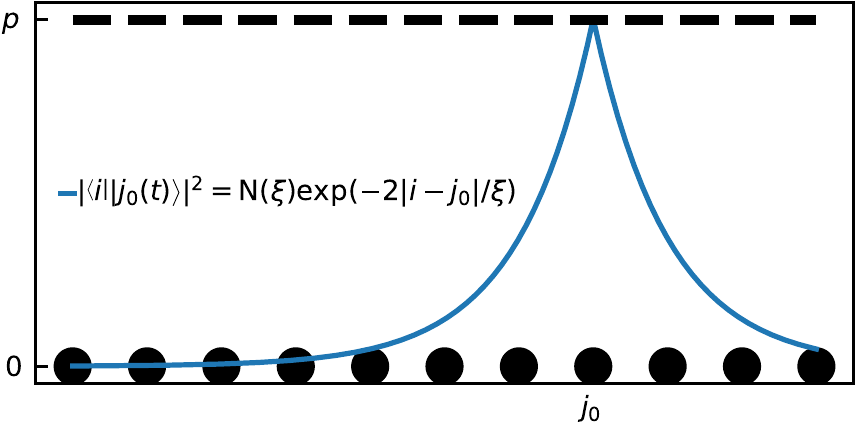}
        \caption{Illustration of the relation between localization length \(\xi\) and waiting probability \(p\): After a click event on site \(j_0\) and successive time evolution over period \(t\) (without further click events), the density \(|\bra{i}\ket{j_0(t)}|^2\) with localization length \(\xi\) establishes. The waiting probability \(p\) to have another click event at \(j_0\) (distance zero jump) is connected to the localization length via the density on the center site \(p = |\bra{j_0} \ket{j_0(t)}|^2 = \mathcal{N}(\xi)\).}
        \label{fig:sketch_x_p_connection}
    \end{figure}

 If there was however a long tail in the distribution \(\mathcal{P}_p(p)\) towards large waiting probabilities \(p \rightarrow 1\), rare regions with atypically large waiting probabilities could slow down the jump-facilitated transport (inducing many distance zero jumps), resulting in subdiffusion. The distribution of waiting probabilities \(\mathcal{P}_p(p)\) is related to the distribution of inverse localization lengths \(y = 1 / \xi\), since the probability for the particle to ``wait" on a site \(\ket{j_0}\) is determined by the peak weight of the wave function \(\bra{i}\ket{j_0(t)}\). We thus make the connection to the localized wave functions, by choosing the distribution of the waiting probabilities \(\mathcal{P}_p(p)\) according to the distribution of inverse localization lengths \(\mathcal{P}_y(y)\)---see Fig. \ref{fig:sketch_x_p_connection} for an illustration.
    
    The probability distribution of the inverse localization length \(y:= 1 / \xi\) of eigenfunctions at a given energy \(E\) and disorder strenth \(W\) is given by a gaussian \cite{Evers-RevModPhys,effective_localization_length},
    \begin{align}
        \mathcal{P}^E_y(y) &= \mathcal{N}_1 \exp\left[-\frac{1}{2}\left(\frac{y - \mu(W, E)}{\sigma(W, E)}\right)^2\right] \qquad y \in [0, \infty), \label{eq:gaussian_localization_length}
    \end{align}
    with normalization constant \(\mathcal{N}_1\), variance \(\sigma^2\), and mean \(\mu\). 
    For the toy model, we approximate the probability to have a zero distance jump \(p_i\) to be given by the center-site maximum of the localized probability density at time \(t\) after the click event
    \begin{align}
        &|\bra{i}\ket{j_0(t)}|^2 = \mathcal{N}(\xi) \exp(-2|i - j_0| / \xi)\\
        &\mathcal{N}(\xi) = \frac{\exp(2 / \xi) - 1}{\exp(2 / \xi) + 1} \qquad L \gg 1
    \end{align}
    This approximation provides a mapping between the random variables \(p\) (representing a waiting probability) and \(y\) (representing an inverse localization length). After a click on an arbitrary site \(j_0\) the waiting probability is given by
    \begin{align}
        p = |\bra{j_0} \ket{j_0(t)}|^2
    \end{align}
    And since the probability on the center site is given by \(p = \mathcal{N}(\xi(y)) = \mathcal{N}(1 / y)\), we get
    \begin{align}
        p(y) = \frac{\exp(2y) - 1}{\exp(2y) + 1}.
    \end{align}
    We use this mapping for the change of variables \(y \rightarrow p\) in \(\mathcal{P}_y(y)\). From \(y \in [0, \infty)\), it follows that waiting probabilities between 1 and 0 can be found: \(p \in [0, 1)\). 
    The probability distribution for the \(p_i\) at a given energy takes the form
    \begin{align}
        \mathcal{P}^E_p(p) &=  \frac{\mathcal{N}_1}{1 - p^2} \exp\left\{- \frac{\left[\log(\frac{1+p}{1-p}) - 2\mu\right]^2}{8\sigma^2}\right\}. 
        \label{eq:onsite_probability_prediction}
    \end{align}
    where the energy dependence is encoded in \(\mu\), \(\sigma\), and \(\mathcal{N}_1\)---see Eq. \eqref{eq:gaussian_localization_length}.
    \par
    Since our time evolution protocol includes all eigenstates, we average over the band to obtain the waiting probability distribution
    \begin{align}
        \mathcal{P}_p(p) = \int d{E} \nu(E) \mathcal{P}_p^E(p) 
        \label{eq:average_prob_dist}
    \end{align}
    with the density of states \(\nu\). The correspondence between \(\mathcal{P}_p(p)\) and \(\mathcal{P}_y(y)\) is only approximate, since the wave function in our time evolution protocol is actually a superposition of eigenfunctions. Due to the exponential decay of the eigenfunctions we can however reasonably replace this superposition by the dominantly contributing eigenfunction (this becomes exact in the limit \(\xi \rightarrow 0\)).

    Using the random-walk picture, we can now address the question of diffusion in \(\Delta_{\mu}(t)\) at asymptotic times.
    As mentioned before the asymptotic behaviour of the random walk is governed by the behaviour of \(\mathcal{P}_p\) for \(p \rightarrow 1\), because sites with a waiting probability of almost one provide bottlenecks for the dynamics in the system. Sites with large waiting probabilities correspond to wave functions with small localization lengths. It is known that spatial randomness in a potential can lead to anomalous transport via long tails towards waiting probability one \cite{random_barrier_heights_conductivity}. It was shown in Ref. \cite{qtm_asymptotics}, that a random walk as specified above behaves diffusively, if the distribution of \(\tau(p):=2 / (1-p)\) has a finite mean value. For our distribution, this mean value exists as it can be easily calculated for distribution \eqref{eq:onsite_probability_prediction}. 
    
    Thus, we find \(\Delta_{\mu}(t) \propto N^{1 / 2} \propto t^{1 / 2}\) in the asymptotic limit.

    As explained in Sec. \ref{sec:measurement_delocalization}, this expectation is independent of the performed averages (provided that \textit{any} average is performed) and holds thus true even if only the average over outcomes is taken. In this case however, on intermediate time scales which are not described by the toy model (\(N \gg L\), \(\Delta_{\rm o} \sim \bar{\xi}\)), few sites have a large impact on the dynamics. This can lead to apparent sub- or even superdiffusive dynamics on these time scales. If a fraction of trajectories reaches a site with \(\xi^{\rm eff} > \bar{\xi}\) (\(\xi^{\rm eff} < \bar{\xi}\)), there is an increase (decrease) with time in the effective localization length (see Fig. \ref{fig:complete_outcome_average} for an example). In the random-walk picture, this may be understood as a time-dependent jump distance distribution, which can raise or lower the local exponent of the power law in \(\Delta_{\rm o}\).

    As a last remark on the model, we reconsider the assumption of time independent \(p_i\). In contrast, in our system, the probabilities \(p_i\) change due to no-click events altering the shape of the wave function, and oscillating contributions of different eigenfunctions to the wave function. However, since we found diffusion including ``memory effects", we would find the same result if the \(p_i(n)\) were taken to be completely uncorrelated in time while drawn from the same distribution.
    \par
    Below, we give a brief summary of the other simplifications in the toy model
    \begin{itemize}
        \item We approximate all jumps distances greater than zero by a unit-distance. This is justified because the probability to jump over a distance \(x\) decreases exponentially with \(x / \xi\).
        \item The model is based on the statistics of the eigenstates of the Anderson Hamiltonian. However, the actual wave functions result from time evolution of one-site states. The approximation is justified by the exponential decay of the eigenstates, which implies that only \(\mathcal{O}(\bar{\xi})\) sites overlap significantly with the original site.
        \item The effect of no-click events on the shape of the wave function is not explicitly taken into account. However, we showed that no-click events can be viewed as a correction to the (effective) localization length. We can incorporate corrections to this simplification on a phenomenological level through a modification of the localization length.
    \end{itemize}
    In summary, by introducing a classical random walk toy model of the particle trajectory ensemble, we are able to confirm that the spread of the ensemble behaves diffusively in the long time limit, thus supporting our findings from Sec. \ref{sec:measurement_delocalization}. In the random walk picture, click events facilitate jumps on the lattice, with the average localization length setting the typical jump distance. Since the wave functions fall off exponentially around a center site, jumps of distance zero are most likely. However, analyzing the distribution of localization lengths, we showed that such waiting events do not lead to subdiffusion.\par
    With the insights from the random walk, we take another look at our measurement steering protocols. Turning first to the blind protocol with random measurements, 
    we can identify the average runtime of the protocol with the maximum expected hitting time of a random walk on a connected graph \cite{random_walk_hitting_times}. This is the number of steps which the random walker needs to take on average, to first arrive on the most distant site, and it scales as \(L^2\) for a simple chain \cite{random_walk_hitting_times}. Since we need \(\propto L\) measurements to induce one step, we find \(\langle N\rangle\propto L^3\), as seen in the numerics.
    For the adaptive protocol, the asymptotic behaviour of the waiting probability distribution confirms a linear relation between the system size and the expected number of measurements, since the moments of the waiting times on a site do not diverge. This also implies that there is a gaussian probability distribution of the steering times for sufficiently large system sizes, in accordance with the central limit theorem.
    \vspace{1cm}
    
    \section{Discussion and Conclusions}
    \label{sec:conclusions}
    We have investigated the dynamics of a single particle in a one-dimensional Anderson-localized system, subject to projective on-site measurements. Performing measurements at random locations, we have found that particle trajectories, driven by click events, spread approximately diffusively over the system. At the same time, the wavefunctions remain localized along individual quantum trajectories, but with a modified effective localization length \ref{sec:measurement_delocalization}. These findings suggest that efficient steering protocols can be formulated, moving the particle through the system to a predefined target site within a number of measurements, which is polynomial in the system size. We have demonstrated this in Sec. \ref{sec:steering}. The average spread of localized entities suggests a random walk picture, which we have utilized in Sec. \ref{sec:random_walk} to confirm the long time diffusive nature of the trajectory spread, and to explain the relation between steering time and system size in the steering protocols.

    For future studies, it would be interesting to consider a generalization to a many-body system in a disordered chain, which is time evolved with a similar projective-measurement protocol. First, quantum statistics of measured particles is expected to influence the dynamics of the system. Second, the effect of measurements may induce correlations between the particles, which could mimic a genuine interparticle interaction that leads to dephasing and many-body delocalization transition. In this respect, it is interesting to explore various types of measurement operators, in particular, those involving simultaneous measurement of two or more particles in a given state. The key question here can be formulated as follows: Is it possible to ``measure'' a many-body localization transition in a non-interacting disordered system that would have been Anderson-localized without measurement?  Finally, while the problem of the measurement-induced entanglement transition was widely discussed for free fermionic chains (see Introduction), understanding the effect of Anderson localization on such a transition remains a challenging problem. 
    
    \textit{Note added:} before submitting our manuscript for publication, we became aware of a related preprint, Ref.~\cite{related_preprint}, addressing monitored disordered free fermions at half-filling.
    \section{Acknowledgements}
    Parts of this work were done in close interaction with Sthitadhi Roy. We are grateful to him for useful discussions and insights. We thank J. F. Karcher, A. D. Mirlin, I. V. Poboiko, M. Szyniszewski and O. Yevtushenko for useful discussions. 
    The work was supported by the Deutsche
    Forschungsgemeinschaft (DFG): Project No. 277101999
    -- TRR 183 (Project C01) and Grants No. EG 96/13-1 and No. GO 1405/6-1, the Helmholtz International Fellow
    Award, and the Israel Binational Science Foundation-- National Science Foundation through award
    DMR-2037654.
    
    \bibliography{bibliography.bib}

\begin{thebibliography}{46}%
\makeatletter
\providecommand \@ifxundefined [1]{%
 \@ifx{#1\undefined}
}%
\providecommand \@ifnum [1]{%
 \ifnum #1\expandafter \@firstoftwo
 \else \expandafter \@secondoftwo
 \fi
}%
\providecommand \@ifx [1]{%
 \ifx #1\expandafter \@firstoftwo
 \else \expandafter \@secondoftwo
 \fi
}%
\providecommand \natexlab [1]{#1}%
\providecommand \enquote  [1]{``#1''}%
\providecommand \bibnamefont  [1]{#1}%
\providecommand \bibfnamefont [1]{#1}%
\providecommand \citenamefont [1]{#1}%
\providecommand \href@noop [0]{\@secondoftwo}%
\providecommand \href [0]{\begingroup \@sanitize@url \@href}%
\providecommand \@href[1]{\@@startlink{#1}\@@href}%
\providecommand \@@href[1]{\endgroup#1\@@endlink}%
\providecommand \@sanitize@url [0]{\catcode `\\12\catcode `\$12\catcode
  `\&12\catcode `\#12\catcode `\^12\catcode `\_12\catcode `\%12\relax}%
\providecommand \@@startlink[1]{}%
\providecommand \@@endlink[0]{}%
\providecommand \url  [0]{\begingroup\@sanitize@url \@url }%
\providecommand \@url [1]{\endgroup\@href {#1}{\urlprefix }}%
\providecommand \urlprefix  [0]{URL }%
\providecommand \Eprint [0]{\href }%
\providecommand \doibase [0]{https://doi.org/}%
\providecommand \selectlanguage [0]{\@gobble}%
\providecommand \bibinfo  [0]{\@secondoftwo}%
\providecommand \bibfield  [0]{\@secondoftwo}%
\providecommand \translation [1]{[#1]}%
\providecommand \BibitemOpen [0]{}%
\providecommand \bibitemStop [0]{}%
\providecommand \bibitemNoStop [0]{.\EOS\space}%
\providecommand \EOS [0]{\spacefactor3000\relax}%
\providecommand \BibitemShut  [1]{\csname bibitem#1\endcsname}%
\let\auto@bib@innerbib\@empty
\bibitem [{\citenamefont {Li}\ \emph {et~al.}(2018)\citenamefont {Li},
  \citenamefont {Chen},\ and\ \citenamefont
  {Fisher}}]{Fisher-PhysRevB.98.205136}%
  \BibitemOpen
  \bibfield  {author} {\bibinfo {author} {\bibfnamefont {Y.}~\bibnamefont
  {Li}}, \bibinfo {author} {\bibfnamefont {X.}~\bibnamefont {Chen}},\ and\
  \bibinfo {author} {\bibfnamefont {M.~P.~A.}\ \bibnamefont {Fisher}},\
  }\bibfield  {title} {\bibinfo {title} {Quantum {Z}eno effect and the
  many-body entanglement transition},\ }\href
  {https://doi.org/10.1103/PhysRevB.98.205136} {\bibfield  {journal} {\bibinfo
  {journal} {Phys. Rev. B}\ }\textbf {\bibinfo {volume} {98}},\ \bibinfo
  {pages} {205136} (\bibinfo {year} {2018})}\BibitemShut {NoStop}%
\bibitem [{\citenamefont {Skinner}\ \emph {et~al.}(2019)\citenamefont
  {Skinner}, \citenamefont {Ruhman},\ and\ \citenamefont
  {Nahum}}]{measured_quantum_circuits_1}%
  \BibitemOpen
  \bibfield  {author} {\bibinfo {author} {\bibfnamefont {B.}~\bibnamefont
  {Skinner}}, \bibinfo {author} {\bibfnamefont {J.}~\bibnamefont {Ruhman}},\
  and\ \bibinfo {author} {\bibfnamefont {A.}~\bibnamefont {Nahum}},\ }\bibfield
   {title} {\bibinfo {title} {Measurement-induced phase transitions in the
  dynamics of entanglement},\ }\href
  {https://doi.org/10.1103/PhysRevX.9.031009} {\bibfield  {journal} {\bibinfo
  {journal} {Phys. Rev. X}\ }\textbf {\bibinfo {volume} {9}},\ \bibinfo {pages}
  {031009} (\bibinfo {year} {2019})}\BibitemShut {NoStop}%
\bibitem [{\citenamefont {Li}\ \emph {et~al.}(2019)\citenamefont {Li},
  \citenamefont {Chen},\ and\ \citenamefont
  {Fisher}}]{measurement_entanglement_transition_quantum_circuits}%
  \BibitemOpen
  \bibfield  {author} {\bibinfo {author} {\bibfnamefont {Y.}~\bibnamefont
  {Li}}, \bibinfo {author} {\bibfnamefont {X.}~\bibnamefont {Chen}},\ and\
  \bibinfo {author} {\bibfnamefont {M.~P.~A.}\ \bibnamefont {Fisher}},\
  }\bibfield  {title} {\bibinfo {title} {Measurement-driven entanglement
  transition in hybrid quantum circuits},\ }\href
  {https://doi.org/10.1103/PhysRevB.100.134306} {\bibfield  {journal} {\bibinfo
   {journal} {Phys. Rev. B}\ }\textbf {\bibinfo {volume} {100}},\ \bibinfo
  {pages} {134306} (\bibinfo {year} {2019})}\BibitemShut {NoStop}%
\bibitem [{\citenamefont {Chan}\ \emph {et~al.}(2019)\citenamefont {Chan},
  \citenamefont {Nandkishore}, \citenamefont {Pretko},\ and\ \citenamefont
  {Smith}}]{Chan-PhysRevB.99.224307}%
  \BibitemOpen
  \bibfield  {author} {\bibinfo {author} {\bibfnamefont {A.}~\bibnamefont
  {Chan}}, \bibinfo {author} {\bibfnamefont {R.~M.}\ \bibnamefont
  {Nandkishore}}, \bibinfo {author} {\bibfnamefont {M.}~\bibnamefont
  {Pretko}},\ and\ \bibinfo {author} {\bibfnamefont {G.}~\bibnamefont
  {Smith}},\ }\bibfield  {title} {\bibinfo {title} {Unitary-projective
  entanglement dynamics},\ }\href {https://doi.org/10.1103/PhysRevB.99.224307}
  {\bibfield  {journal} {\bibinfo  {journal} {Phys. Rev. B}\ }\textbf {\bibinfo
  {volume} {99}},\ \bibinfo {pages} {224307} (\bibinfo {year}
  {2019})}\BibitemShut {NoStop}%
\bibitem [{\citenamefont {Bao}\ \emph {et~al.}(2020)\citenamefont {Bao},
  \citenamefont {Choi},\ and\ \citenamefont
  {Altman}}]{Altman-PhysRevB.101.104301}%
  \BibitemOpen
  \bibfield  {author} {\bibinfo {author} {\bibfnamefont {Y.}~\bibnamefont
  {Bao}}, \bibinfo {author} {\bibfnamefont {S.}~\bibnamefont {Choi}},\ and\
  \bibinfo {author} {\bibfnamefont {E.}~\bibnamefont {Altman}},\ }\bibfield
  {title} {\bibinfo {title} {Theory of the phase transition in random unitary
  circuits with measurements},\ }\href
  {https://doi.org/10.1103/PhysRevB.101.104301} {\bibfield  {journal} {\bibinfo
   {journal} {Phys. Rev. B}\ }\textbf {\bibinfo {volume} {101}},\ \bibinfo
  {pages} {104301} (\bibinfo {year} {2020})}\BibitemShut {NoStop}%
\bibitem [{\citenamefont {Szyniszewski}\ \emph {et~al.}(2019)\citenamefont
  {Szyniszewski}, \citenamefont {Romito},\ and\ \citenamefont
  {Schomerus}}]{Romito-PhysRevB.100.064204}%
  \BibitemOpen
  \bibfield  {author} {\bibinfo {author} {\bibfnamefont {M.}~\bibnamefont
  {Szyniszewski}}, \bibinfo {author} {\bibfnamefont {A.}~\bibnamefont
  {Romito}},\ and\ \bibinfo {author} {\bibfnamefont {H.}~\bibnamefont
  {Schomerus}},\ }\bibfield  {title} {\bibinfo {title} {Entanglement transition
  from variable-strength weak measurements},\ }\href
  {https://doi.org/10.1103/PhysRevB.100.064204} {\bibfield  {journal} {\bibinfo
   {journal} {Phys. Rev. B}\ }\textbf {\bibinfo {volume} {100}},\ \bibinfo
  {pages} {064204} (\bibinfo {year} {2019})}\BibitemShut {NoStop}%
\bibitem [{\citenamefont {Fisher}\ \emph {et~al.}(2022)\citenamefont {Fisher},
  \citenamefont {Khemani}, \citenamefont {Nahum},\ and\ \citenamefont
  {Vijay}}]{quantum_circuits_review}%
  \BibitemOpen
  \bibfield  {author} {\bibinfo {author} {\bibfnamefont {M.~P.~A.}\
  \bibnamefont {Fisher}}, \bibinfo {author} {\bibfnamefont {V.}~\bibnamefont
  {Khemani}}, \bibinfo {author} {\bibfnamefont {A.}~\bibnamefont {Nahum}},\
  and\ \bibinfo {author} {\bibfnamefont {S.}~\bibnamefont {Vijay}},\ }\href
  {https://doi.org/10.48550/ARXIV.2207.14280} {\bibinfo {title} {Random quantum
  circuits}} (\bibinfo {year} {2022}),\ \bibinfo {note}
  {arXiv:2207.14280}\BibitemShut {NoStop}%
\bibitem [{\citenamefont {Alberton}\ \emph {et~al.}(2021)\citenamefont
  {Alberton}, \citenamefont {Buchhold},\ and\ \citenamefont
  {Diehl}}]{entanglement_transition_monitored_free_fermion_chain}%
  \BibitemOpen
  \bibfield  {author} {\bibinfo {author} {\bibfnamefont {O.}~\bibnamefont
  {Alberton}}, \bibinfo {author} {\bibfnamefont {M.}~\bibnamefont {Buchhold}},\
  and\ \bibinfo {author} {\bibfnamefont {S.}~\bibnamefont {Diehl}},\ }\bibfield
   {title} {\bibinfo {title} {Entanglement transition in a monitored
  free-fermion chain: {F}rom extended criticality to area law},\ }\href
  {https://doi.org/10.1103/PhysRevLett.126.170602} {\bibfield  {journal}
  {\bibinfo  {journal} {Phys. Rev. Lett.}\ }\textbf {\bibinfo {volume} {126}},\
  \bibinfo {pages} {170602} (\bibinfo {year} {2021})}\BibitemShut {NoStop}%
\bibitem [{\citenamefont {Buchhold}\ \emph {et~al.}(2021)\citenamefont
  {Buchhold}, \citenamefont {Minoguchi}, \citenamefont {Altland},\ and\
  \citenamefont {Diehl}}]{measured_dirac_fermions}%
  \BibitemOpen
  \bibfield  {author} {\bibinfo {author} {\bibfnamefont {M.}~\bibnamefont
  {Buchhold}}, \bibinfo {author} {\bibfnamefont {Y.}~\bibnamefont {Minoguchi}},
  \bibinfo {author} {\bibfnamefont {A.}~\bibnamefont {Altland}},\ and\ \bibinfo
  {author} {\bibfnamefont {S.}~\bibnamefont {Diehl}},\ }\bibfield  {title}
  {\bibinfo {title} {Effective theory for the measurement-induced phase
  transition of {D}irac fermions},\ }\href
  {https://doi.org/10.1103/PhysRevX.11.041004} {\bibfield  {journal} {\bibinfo
  {journal} {Phys. Rev. X}\ }\textbf {\bibinfo {volume} {11}},\ \bibinfo
  {pages} {041004} (\bibinfo {year} {2021})}\BibitemShut {NoStop}%
\bibitem [{\citenamefont {Doggen}\ \emph {et~al.}(2022)\citenamefont {Doggen},
  \citenamefont {Gefen}, \citenamefont {Gornyi}, \citenamefont {Mirlin},\ and\
  \citenamefont {Polyakov}}]{DGG_measured_chain}%
  \BibitemOpen
  \bibfield  {author} {\bibinfo {author} {\bibfnamefont {E.~V.~H.}\
  \bibnamefont {Doggen}}, \bibinfo {author} {\bibfnamefont {Y.}~\bibnamefont
  {Gefen}}, \bibinfo {author} {\bibfnamefont {I.~V.}\ \bibnamefont {Gornyi}},
  \bibinfo {author} {\bibfnamefont {A.~D.}\ \bibnamefont {Mirlin}},\ and\
  \bibinfo {author} {\bibfnamefont {D.~G.}\ \bibnamefont {Polyakov}},\
  }\bibfield  {title} {\bibinfo {title} {Generalized quantum measurements with
  matrix product states: {E}ntanglement phase transition and clusterization},\
  }\href {https://doi.org/10.1103/PhysRevResearch.4.023146} {\bibfield
  {journal} {\bibinfo  {journal} {Phys. Rev. Research}\ }\textbf {\bibinfo
  {volume} {4}},\ \bibinfo {pages} {023146} (\bibinfo {year}
  {2022})}\BibitemShut {NoStop}%
\bibitem [{\citenamefont {Merritt}\ and\ \citenamefont
  {Fidkowski}(2022)}]{entanglement_transitions_with_free_fermions}%
  \BibitemOpen
  \bibfield  {author} {\bibinfo {author} {\bibfnamefont {J.}~\bibnamefont
  {Merritt}}\ and\ \bibinfo {author} {\bibfnamefont {L.}~\bibnamefont
  {Fidkowski}},\ }\href {https://doi.org/10.48550/ARXIV.2210.05681} {\bibinfo
  {title} {Entanglement transitions with free fermions}} (\bibinfo {year}
  {2022}),\ \bibinfo {note} {arXiv:2210.05681}\BibitemShut {NoStop}%
\bibitem [{\citenamefont {Cao}\ \emph {et~al.}(2019)\citenamefont {Cao},
  \citenamefont {Tilloy},\ and\ \citenamefont
  {Luca}}]{entanglement_continuously_mintored_chain}%
  \BibitemOpen
  \bibfield  {author} {\bibinfo {author} {\bibfnamefont {X.}~\bibnamefont
  {Cao}}, \bibinfo {author} {\bibfnamefont {A.}~\bibnamefont {Tilloy}},\ and\
  \bibinfo {author} {\bibfnamefont {A.~D.}\ \bibnamefont {Luca}},\ }\bibfield
  {title} {\bibinfo {title} {{Entanglement in a fermion chain under continuous
  monitoring}},\ }\href {https://doi.org/10.21468/SciPostPhys.7.2.024}
  {\bibfield  {journal} {\bibinfo  {journal} {SciPost Phys.}\ }\textbf
  {\bibinfo {volume} {7}},\ \bibinfo {pages} {024} (\bibinfo {year}
  {2019})}\BibitemShut {NoStop}%
\bibitem [{\citenamefont {Turkeshi}\ \emph {et~al.}(2022)\citenamefont
  {Turkeshi}, \citenamefont {Dalmonte}, \citenamefont {Fazio},\ and\
  \citenamefont {Schir\`o}}]{entangelement_transitions_quasiparticles}%
  \BibitemOpen
  \bibfield  {author} {\bibinfo {author} {\bibfnamefont {X.}~\bibnamefont
  {Turkeshi}}, \bibinfo {author} {\bibfnamefont {M.}~\bibnamefont {Dalmonte}},
  \bibinfo {author} {\bibfnamefont {R.}~\bibnamefont {Fazio}},\ and\ \bibinfo
  {author} {\bibfnamefont {M.}~\bibnamefont {Schir\`o}},\ }\bibfield  {title}
  {\bibinfo {title} {Entanglement transitions from stochastic resetting of
  non-{H}ermitian quasiparticles},\ }\href
  {https://doi.org/10.1103/PhysRevB.105.L241114} {\bibfield  {journal}
  {\bibinfo  {journal} {Phys. Rev. B}\ }\textbf {\bibinfo {volume} {105}},\
  \bibinfo {pages} {L241114} (\bibinfo {year} {2022})}\BibitemShut {NoStop}%
\bibitem [{\citenamefont {Altland}\ \emph {et~al.}(2022)\citenamefont
  {Altland}, \citenamefont {Buchhold}, \citenamefont {Diehl},\ and\
  \citenamefont {Micklitz}}]{dynamics_of_measured_mb_chaotic_systems}%
  \BibitemOpen
  \bibfield  {author} {\bibinfo {author} {\bibfnamefont {A.}~\bibnamefont
  {Altland}}, \bibinfo {author} {\bibfnamefont {M.}~\bibnamefont {Buchhold}},
  \bibinfo {author} {\bibfnamefont {S.}~\bibnamefont {Diehl}},\ and\ \bibinfo
  {author} {\bibfnamefont {T.}~\bibnamefont {Micklitz}},\ }\bibfield  {title}
  {\bibinfo {title} {Dynamics of measured many-body quantum chaotic systems},\
  }\href {https://doi.org/10.1103/PhysRevResearch.4.L022066} {\bibfield
  {journal} {\bibinfo  {journal} {Phys. Rev. Research}\ }\textbf {\bibinfo
  {volume} {4}},\ \bibinfo {pages} {L022066} (\bibinfo {year}
  {2022})}\BibitemShut {NoStop}%
\bibitem [{\citenamefont {Ladewig}\ \emph {et~al.}(2022)\citenamefont
  {Ladewig}, \citenamefont {Diehl},\ and\ \citenamefont
  {Buchhold}}]{interplay_measurement_decoherence_free_hamiltonian_evolution}%
  \BibitemOpen
  \bibfield  {author} {\bibinfo {author} {\bibfnamefont {B.}~\bibnamefont
  {Ladewig}}, \bibinfo {author} {\bibfnamefont {S.}~\bibnamefont {Diehl}},\
  and\ \bibinfo {author} {\bibfnamefont {M.}~\bibnamefont {Buchhold}},\
  }\bibfield  {title} {\bibinfo {title} {Monitored open fermion dynamics:
  {E}xploring the interplay of measurement, decoherence, and free {H}amiltonian
  evolution},\ }\href {https://doi.org/10.1103/PhysRevResearch.4.033001}
  {\bibfield  {journal} {\bibinfo  {journal} {Phys. Rev. Research}\ }\textbf
  {\bibinfo {volume} {4}},\ \bibinfo {pages} {033001} (\bibinfo {year}
  {2022})}\BibitemShut {NoStop}%
\bibitem [{\citenamefont {Minoguchi}\ \emph {et~al.}(2022)\citenamefont
  {Minoguchi}, \citenamefont {Rabl},\ and\ \citenamefont
  {Buchhold}}]{continuous_measurement_free_bosons}%
  \BibitemOpen
  \bibfield  {author} {\bibinfo {author} {\bibfnamefont {Y.}~\bibnamefont
  {Minoguchi}}, \bibinfo {author} {\bibfnamefont {P.}~\bibnamefont {Rabl}},\
  and\ \bibinfo {author} {\bibfnamefont {M.}~\bibnamefont {Buchhold}},\
  }\bibfield  {title} {\bibinfo {title} {{Continuous {G}aussian measurements of
  the free boson {CFT}: {A} model for exactly solvable and detectable
  measurement-induced dynamics}},\ }\href
  {https://doi.org/10.21468/SciPostPhys.12.1.009} {\bibfield  {journal}
  {\bibinfo  {journal} {SciPost Phys.}\ }\textbf {\bibinfo {volume} {12}},\
  \bibinfo {pages} {009} (\bibinfo {year} {2022})}\BibitemShut {NoStop}%
\bibitem [{\citenamefont {Yang}\ \emph {et~al.}(2022)\citenamefont {Yang},
  \citenamefont {Zuo},\ and\ \citenamefont
  {Liu}}]{continuously_measured_free_fermion_gas}%
  \BibitemOpen
  \bibfield  {author} {\bibinfo {author} {\bibfnamefont {Q.}~\bibnamefont
  {Yang}}, \bibinfo {author} {\bibfnamefont {Y.}~\bibnamefont {Zuo}},\ and\
  \bibinfo {author} {\bibfnamefont {D.~E.}\ \bibnamefont {Liu}},\ }\href
  {https://doi.org/10.48550/ARXIV.2207.03376} {\bibinfo {title} {Keldysh
  nonlinear sigma model for a free-fermion gas under continuous measurements}}
  (\bibinfo {year} {2022}),\ \bibinfo {note} {arXiv:2207.03376}\BibitemShut
  {NoStop}%
\bibitem [{\citenamefont {Dhar}\ \emph
  {et~al.}(2015{\natexlab{a}})\citenamefont {Dhar}, \citenamefont {Dasgupta},\
  and\ \citenamefont {Dhar}}]{quantum_time_of_arrival}%
  \BibitemOpen
  \bibfield  {author} {\bibinfo {author} {\bibfnamefont {S.}~\bibnamefont
  {Dhar}}, \bibinfo {author} {\bibfnamefont {S.}~\bibnamefont {Dasgupta}},\
  and\ \bibinfo {author} {\bibfnamefont {A.}~\bibnamefont {Dhar}},\ }\bibfield
  {title} {\bibinfo {title} {Quantum time of arrival distribution in a simple
  lattice model},\ }\href {https://doi.org/10.1088/1751-8113/48/11/115304}
  {\bibfield  {journal} {\bibinfo  {journal} {Journal of Physics A:
  Mathematical and Theoretical}\ }\textbf {\bibinfo {volume} {48}},\ \bibinfo
  {pages} {115304} (\bibinfo {year} {2015}{\natexlab{a}})}\BibitemShut
  {NoStop}%
\bibitem [{\citenamefont {Dhar}\ \emph
  {et~al.}(2015{\natexlab{b}})\citenamefont {Dhar}, \citenamefont {Dasgupta},
  \citenamefont {Dhar},\ and\ \citenamefont
  {Sen}}]{projective_measurements_tb}%
  \BibitemOpen
  \bibfield  {author} {\bibinfo {author} {\bibfnamefont {S.}~\bibnamefont
  {Dhar}}, \bibinfo {author} {\bibfnamefont {S.}~\bibnamefont {Dasgupta}},
  \bibinfo {author} {\bibfnamefont {A.}~\bibnamefont {Dhar}},\ and\ \bibinfo
  {author} {\bibfnamefont {D.}~\bibnamefont {Sen}},\ }\bibfield  {title}
  {\bibinfo {title} {Detection of a quantum particle on a lattice under
  repeated projective measurements},\ }\href
  {https://doi.org/10.1103/PhysRevA.91.062115} {\bibfield  {journal} {\bibinfo
  {journal} {Phys. Rev. A}\ }\textbf {\bibinfo {volume} {91}},\ \bibinfo
  {pages} {062115} (\bibinfo {year} {2015}{\natexlab{b}})}\BibitemShut
  {NoStop}%
\bibitem [{\citenamefont
  {Alba}(2022)}]{entanglement_production_dissipative_impurity}%
  \BibitemOpen
  \bibfield  {author} {\bibinfo {author} {\bibfnamefont {V.}~\bibnamefont
  {Alba}},\ }\bibfield  {title} {\bibinfo {title} {{Unbounded entanglement
  production via a dissipative impurity}},\ }\href
  {https://doi.org/10.21468/SciPostPhys.12.1.011} {\bibfield  {journal}
  {\bibinfo  {journal} {SciPost Phys.}\ }\textbf {\bibinfo {volume} {12}},\
  \bibinfo {pages} {011} (\bibinfo {year} {2022})}\BibitemShut {NoStop}%
\bibitem [{\citenamefont {Caceffo}\ and\ \citenamefont
  {Alba}(2022)}]{fermionic_chain_with_dissipative_defects}%
  \BibitemOpen
  \bibfield  {author} {\bibinfo {author} {\bibfnamefont {F.}~\bibnamefont
  {Caceffo}}\ and\ \bibinfo {author} {\bibfnamefont {V.}~\bibnamefont {Alba}},\
  }\href {https://doi.org/10.48550/ARXIV.2209.14164} {\bibinfo {title}
  {Entanglement negativity in a fermionic chain with dissipative defects:
  {E}xact results}} (\bibinfo {year} {2022}),\ \bibinfo {note}
  {arXiv:2209.14164}\BibitemShut {NoStop}%
\bibitem [{\citenamefont {Jin}\ \emph {et~al.}(2022)\citenamefont {Jin},
  \citenamefont {Ferreira}, \citenamefont {Bauer}, \citenamefont {Filippone},\
  and\ \citenamefont {Giamarchi}}]{stochastic_resistors}%
  \BibitemOpen
  \bibfield  {author} {\bibinfo {author} {\bibfnamefont {T.}~\bibnamefont
  {Jin}}, \bibinfo {author} {\bibfnamefont {J.}~\bibnamefont {Ferreira}},
  \bibinfo {author} {\bibfnamefont {M.}~\bibnamefont {Bauer}}, \bibinfo
  {author} {\bibfnamefont {M.}~\bibnamefont {Filippone}},\ and\ \bibinfo
  {author} {\bibfnamefont {T.}~\bibnamefont {Giamarchi}},\ }\href
  {https://doi.org/10.48550/ARXIV.2206.13985} {\bibinfo {title} {Semi-classical
  theory of quantum stochastic resistors}} (\bibinfo {year} {2022}),\ \bibinfo
  {note} {arXiv:2206.13985}\BibitemShut {NoStop}%
\bibitem [{\citenamefont {Anderson}(1958)}]{paper:absence_of_diffusion}%
  \BibitemOpen
  \bibfield  {author} {\bibinfo {author} {\bibfnamefont {P.~W.}\ \bibnamefont
  {Anderson}},\ }\bibfield  {title} {\bibinfo {title} {{Absence of Diffusion in
  Certain Random Lattices}},\ }\href {https://doi.org/10.1103/PhysRev.109.1492}
  {\bibfield  {journal} {\bibinfo  {journal} {Phys. Rev.}\ }\textbf {\bibinfo
  {volume} {109}},\ \bibinfo {pages} {1492} (\bibinfo {year}
  {1958})}\BibitemShut {NoStop}%
\bibitem [{\citenamefont {Abrahams}\ \emph {et~al.}(1979)\citenamefont
  {Abrahams}, \citenamefont {Anderson}, \citenamefont {Licciardello},\ and\
  \citenamefont {Ramakrishnan}}]{paper:localization_in_two_d}%
  \BibitemOpen
  \bibfield  {author} {\bibinfo {author} {\bibfnamefont {E.}~\bibnamefont
  {Abrahams}}, \bibinfo {author} {\bibfnamefont {P.~W.}\ \bibnamefont
  {Anderson}}, \bibinfo {author} {\bibfnamefont {D.~C.}\ \bibnamefont
  {Licciardello}},\ and\ \bibinfo {author} {\bibfnamefont {T.~V.}\ \bibnamefont
  {Ramakrishnan}},\ }\bibfield  {title} {\bibinfo {title} {Scaling theory of
  localization: {A}bsence of quantum diffusion in two dimensions},\ }\href
  {https://doi.org/10.1103/PhysRevLett.42.673} {\bibfield  {journal} {\bibinfo
  {journal} {Phys. Rev. Lett.}\ }\textbf {\bibinfo {volume} {42}},\ \bibinfo
  {pages} {673} (\bibinfo {year} {1979})}\BibitemShut {NoStop}%
\bibitem [{\citenamefont {Evers}\ and\ \citenamefont
  {Mirlin}(2008)}]{Evers-RevModPhys}%
  \BibitemOpen
  \bibfield  {author} {\bibinfo {author} {\bibfnamefont {F.}~\bibnamefont
  {Evers}}\ and\ \bibinfo {author} {\bibfnamefont {A.~D.}\ \bibnamefont
  {Mirlin}},\ }\bibfield  {title} {\bibinfo {title} {Anderson transitions},\
  }\href {https://doi.org/10.1103/RevModPhys.80.1355} {\bibfield  {journal}
  {\bibinfo  {journal} {Rev. Mod. Phys.}\ }\textbf {\bibinfo {volume} {80}},\
  \bibinfo {pages} {1355} (\bibinfo {year} {2008})}\BibitemShut {NoStop}%
\bibitem [{\citenamefont {Gopalakrishnan}\ \emph {et~al.}(2017)\citenamefont
  {Gopalakrishnan}, \citenamefont {Islam},\ and\ \citenamefont
  {Knap}}]{gopalakrishnan_knap_noise_and_diffusion}%
  \BibitemOpen
  \bibfield  {author} {\bibinfo {author} {\bibfnamefont {S.}~\bibnamefont
  {Gopalakrishnan}}, \bibinfo {author} {\bibfnamefont {K.~R.}\ \bibnamefont
  {Islam}},\ and\ \bibinfo {author} {\bibfnamefont {M.}~\bibnamefont {Knap}},\
  }\bibfield  {title} {\bibinfo {title} {Noise-induced subdiffusion in strongly
  localized quantum systems},\ }\href
  {https://doi.org/10.1103/PhysRevLett.119.046601} {\bibfield  {journal}
  {\bibinfo  {journal} {Phys. Rev. Lett.}\ }\textbf {\bibinfo {volume} {119}},\
  \bibinfo {pages} {046601} (\bibinfo {year} {2017})}\BibitemShut {NoStop}%
\bibitem [{\citenamefont {Lunt}\ and\ \citenamefont
  {Pal}(2020)}]{mbl_measurements}%
  \BibitemOpen
  \bibfield  {author} {\bibinfo {author} {\bibfnamefont {O.}~\bibnamefont
  {Lunt}}\ and\ \bibinfo {author} {\bibfnamefont {A.}~\bibnamefont {Pal}},\
  }\bibfield  {title} {\bibinfo {title} {Measurement-induced entanglement
  transitions in many-body localized systems},\ }\href
  {https://doi.org/10.1103/PhysRevResearch.2.043072} {\bibfield  {journal}
  {\bibinfo  {journal} {Phys. Rev. Research}\ }\textbf {\bibinfo {volume}
  {2}},\ \bibinfo {pages} {043072} (\bibinfo {year} {2020})}\BibitemShut
  {NoStop}%
\bibitem [{\citenamefont {Roy}\ \emph {et~al.}(2020)\citenamefont {Roy},
  \citenamefont {Chalker}, \citenamefont {Gornyi},\ and\ \citenamefont
  {Gefen}}]{steering_rcgg}%
  \BibitemOpen
  \bibfield  {author} {\bibinfo {author} {\bibfnamefont {S.}~\bibnamefont
  {Roy}}, \bibinfo {author} {\bibfnamefont {J.~T.}\ \bibnamefont {Chalker}},
  \bibinfo {author} {\bibfnamefont {I.~V.}\ \bibnamefont {Gornyi}},\ and\
  \bibinfo {author} {\bibfnamefont {Y.}~\bibnamefont {Gefen}},\ }\bibfield
  {title} {\bibinfo {title} {Measurement-induced steering of quantum systems},\
  }\href {https://doi.org/10.1103/PhysRevResearch.2.033347} {\bibfield
  {journal} {\bibinfo  {journal} {Phys. Rev. Research}\ }\textbf {\bibinfo
  {volume} {2}},\ \bibinfo {pages} {033347} (\bibinfo {year}
  {2020})}\BibitemShut {NoStop}%
\bibitem [{\citenamefont {Herasymenko}\ \emph {et~al.}(2021)\citenamefont
  {Herasymenko}, \citenamefont {Gornyi},\ and\ \citenamefont
  {Gefen}}]{Herasymenko}%
  \BibitemOpen
  \bibfield  {author} {\bibinfo {author} {\bibfnamefont {Y.}~\bibnamefont
  {Herasymenko}}, \bibinfo {author} {\bibfnamefont {I.}~\bibnamefont
  {Gornyi}},\ and\ \bibinfo {author} {\bibfnamefont {Y.}~\bibnamefont
  {Gefen}},\ }\href {https://doi.org/10.48550/ARXIV.2111.09306} {\bibinfo
  {title} {Measurement-driven navigation in many-body hilbert space:
  {A}ctive-decision steering}} (\bibinfo {year} {2021}),\ \bibinfo {note}
  {arXiv:2111.09306}\BibitemShut {NoStop}%
\bibitem [{\citenamefont {Garratt}\ \emph {et~al.}(2022)\citenamefont
  {Garratt}, \citenamefont {Weinstein},\ and\ \citenamefont
  {Altman}}]{measurements_restructure_critical_states}%
  \BibitemOpen
  \bibfield  {author} {\bibinfo {author} {\bibfnamefont {S.~J.}\ \bibnamefont
  {Garratt}}, \bibinfo {author} {\bibfnamefont {Z.}~\bibnamefont {Weinstein}},\
  and\ \bibinfo {author} {\bibfnamefont {E.}~\bibnamefont {Altman}},\ }\href
  {https://doi.org/10.48550/ARXIV.2207.09476} {\bibinfo {title} {Measurements
  conspire nonlocally to restructure critical quantum states}} (\bibinfo {year}
  {2022}),\ \bibinfo {note} {arXiv:2207.09476}\BibitemShut {NoStop}%
\bibitem [{\citenamefont {Wampler}\ \emph {et~al.}(2022)\citenamefont
  {Wampler}, \citenamefont {Khor}, \citenamefont {Refael},\ and\ \citenamefont
  {Klich}}]{Klich-PhysRevX.12.031031}%
  \BibitemOpen
  \bibfield  {author} {\bibinfo {author} {\bibfnamefont {M.}~\bibnamefont
  {Wampler}}, \bibinfo {author} {\bibfnamefont {B.~J.~J.}\ \bibnamefont
  {Khor}}, \bibinfo {author} {\bibfnamefont {G.}~\bibnamefont {Refael}},\ and\
  \bibinfo {author} {\bibfnamefont {I.}~\bibnamefont {Klich}},\ }\bibfield
  {title} {\bibinfo {title} {Stirring by staring: {M}easurement-induced
  chirality},\ }\href {https://doi.org/10.1103/PhysRevX.12.031031} {\bibfield
  {journal} {\bibinfo  {journal} {Phys. Rev. X}\ }\textbf {\bibinfo {volume}
  {12}},\ \bibinfo {pages} {031031} (\bibinfo {year} {2022})}\BibitemShut
  {NoStop}%
\bibitem [{\citenamefont {Misra}\ and\ \citenamefont
  {Sudarshan}(1977)}]{Misra1977}%
  \BibitemOpen
  \bibfield  {author} {\bibinfo {author} {\bibfnamefont {B.}~\bibnamefont
  {Misra}}\ and\ \bibinfo {author} {\bibfnamefont {E.~C.~G.}\ \bibnamefont
  {Sudarshan}},\ }\bibfield  {title} {\bibinfo {title} {The {Z}eno’s paradox
  in quantum theory},\ }\href {https://doi.org/10.1063/1.523304} {\bibfield
  {journal} {\bibinfo  {journal} {Journal of Mathematical Physics}\ }\textbf
  {\bibinfo {volume} {18}},\ \bibinfo {pages} {756} (\bibinfo {year} {1977})},\
  \Eprint {https://arxiv.org/abs/https://doi.org/10.1063/1.523304}
  {https://doi.org/10.1063/1.523304} \BibitemShut {NoStop}%
\bibitem [{\citenamefont {Peres}(1980)}]{Peres1980}%
  \BibitemOpen
  \bibfield  {author} {\bibinfo {author} {\bibfnamefont {A.}~\bibnamefont
  {Peres}},\ }\bibfield  {title} {\bibinfo {title} {Zeno paradox in quantum
  theory},\ }\href {https://doi.org/10.1119/1.12204} {\bibfield  {journal}
  {\bibinfo  {journal} {American Journal of Physics}\ }\textbf {\bibinfo
  {volume} {48}},\ \bibinfo {pages} {931} (\bibinfo {year} {1980})},\ \Eprint
  {https://arxiv.org/abs/https://doi.org/10.1119/1.12204}
  {https://doi.org/10.1119/1.12204} \BibitemShut {NoStop}%
\bibitem [{\citenamefont {Chaudhry}(2016)}]{Chaudhry2016}%
  \BibitemOpen
  \bibfield  {author} {\bibinfo {author} {\bibfnamefont {A.~Z.}\ \bibnamefont
  {Chaudhry}},\ }\bibfield  {title} {\bibinfo {title} {A general framework for
  the quantum {Z}eno and anti-{Z}eno effects},\ }\href
  {https://doi.org/10.1038/srep29497} {\bibfield  {journal} {\bibinfo
  {journal} {Scientific Reports}\ }\textbf {\bibinfo {volume} {6}},\ \bibinfo
  {pages} {2045} (\bibinfo {year} {2016})},\ \Eprint
  {https://arxiv.org/abs/https://doi.org/10.1038/srep29497}
  {https://doi.org/10.1038/srep29497} \BibitemShut {NoStop}%
\bibitem [{\citenamefont {Snizhko}\ \emph {et~al.}(2020)\citenamefont
  {Snizhko}, \citenamefont {Kumar},\ and\ \citenamefont
  {Romito}}]{Snizhko2020}%
  \BibitemOpen
  \bibfield  {author} {\bibinfo {author} {\bibfnamefont {K.}~\bibnamefont
  {Snizhko}}, \bibinfo {author} {\bibfnamefont {P.}~\bibnamefont {Kumar}},\
  and\ \bibinfo {author} {\bibfnamefont {A.}~\bibnamefont {Romito}},\
  }\bibfield  {title} {\bibinfo {title} {Quantum {Z}eno effect appears in
  stages},\ }\href {https://doi.org/10.1103/PhysRevResearch.2.033512}
  {\bibfield  {journal} {\bibinfo  {journal} {Phys. Rev. Res.}\ }\textbf
  {\bibinfo {volume} {2}},\ \bibinfo {pages} {033512} (\bibinfo {year}
  {2020})}\BibitemShut {NoStop}%
\bibitem [{\citenamefont {Didi}\ and\ \citenamefont
  {Barkai}(2022)}]{didi2021measurement}%
  \BibitemOpen
  \bibfield  {author} {\bibinfo {author} {\bibfnamefont {A.}~\bibnamefont
  {Didi}}\ and\ \bibinfo {author} {\bibfnamefont {E.}~\bibnamefont {Barkai}},\
  }\bibfield  {title} {\bibinfo {title} {Measurement-induced quantum walks},\
  }\href {https://doi.org/10.1103/PhysRevE.105.054108} {\bibfield  {journal}
  {\bibinfo  {journal} {Phys. Rev. E}\ }\textbf {\bibinfo {volume} {105}},\
  \bibinfo {pages} {054108} (\bibinfo {year} {2022})}\BibitemShut {NoStop}%
\bibitem [{\citenamefont {Perrin}\ \emph {et~al.}(2022)\citenamefont {Perrin},
  \citenamefont {Fuchs},\ and\ \citenamefont {Mosseri}}]{hugo_random_walks}%
  \BibitemOpen
  \bibfield  {author} {\bibinfo {author} {\bibfnamefont {H.}~\bibnamefont
  {Perrin}}, \bibinfo {author} {\bibfnamefont {J.-N.}\ \bibnamefont {Fuchs}},\
  and\ \bibinfo {author} {\bibfnamefont {R.}~\bibnamefont {Mosseri}},\
  }\bibfield  {title} {\bibinfo {title} {Robustness of {A}haronov-{B}ohm cages
  in quantum walks},\ }\href {https://doi.org/10.1103/PhysRevB.105.235404}
  {\bibfield  {journal} {\bibinfo  {journal} {Phys. Rev. B}\ }\textbf {\bibinfo
  {volume} {105}},\ \bibinfo {pages} {235404} (\bibinfo {year}
  {2022})}\BibitemShut {NoStop}%
\bibitem [{\citenamefont {Wischmann}\ and\ \citenamefont
  {M{\"u}ller-Hartmann}(1990)}]{effective_localization_length}%
  \BibitemOpen
  \bibfield  {author} {\bibinfo {author} {\bibfnamefont {B.}~\bibnamefont
  {Wischmann}}\ and\ \bibinfo {author} {\bibfnamefont {E.}~\bibnamefont
  {M{\"u}ller-Hartmann}},\ }\bibfield  {title} {\bibinfo {title} {{Level
  statistics and localization: A study of the 1D Anderson model}},\ }\href@noop
  {} {\bibfield  {journal} {\bibinfo  {journal} {Zeitschrift f{\"u}r Physik B
  Condensed Matter}\ }\textbf {\bibinfo {volume} {79}},\ \bibinfo {pages} {91}
  (\bibinfo {year} {1990})}\BibitemShut {NoStop}%
\bibitem [{Note1()}]{Note1}%
  \BibitemOpen
  \bibinfo {note} {This relation holds true if \(r_1\) is calculated as in a
  system with open boundary conditions, according to Eq.~\protect \textup
  {\hbox {\mathsurround \z@ \protect \normalfont (\ignorespaces \ref
  {eq:rq}\unskip \@@italiccorr )}}. Using periodic boundary conditions, \(r_1\)
  needs to be defined more carefully~\cite {com_periodic_bcs} to give correct
  values for wave functions close to the boundaries. In general, as long as the
  trajectory spread is much smaller than the system size, the boundary
  conditions make no difference, but if there are many trajectories close to
  periodic ``boundaries'', the definitions of all observables must be adapted
  carefully. Since the numerical results shown in Sec. \ref
  {sec:measurement_delocalization} were obtained with periodic boundary
  conditions, we exercised caution. In most cases, the shift of the wave
  function is much smaller than the system size such the given definitions can
  be employed. When we observe saturation of \(\Delta \), we calculate the
  position average as described in Ref. \cite {com_periodic_bcs}, and obtain
  \(\xi ^{\protect \rm eff}\) by first shifting the center of the wave function
  to the center site of the system. In this case, \(\Delta ^{\protect \rm
  class}\) should be explicitly calculated from \(\langle r_1^2\rangle -
  \langle r_1\rangle ^2\)}\BibitemShut {NoStop}%
\bibitem [{\citenamefont {Jian}\ \emph {et~al.}(2023)\citenamefont {Jian},
  \citenamefont {Shapourian}, \citenamefont {Bauer},\ and\ \citenamefont
  {Ludwig}}]{forced_measurement}%
  \BibitemOpen
  \bibfield  {author} {\bibinfo {author} {\bibfnamefont {C.-M.}\ \bibnamefont
  {Jian}}, \bibinfo {author} {\bibfnamefont {H.}~\bibnamefont {Shapourian}},
  \bibinfo {author} {\bibfnamefont {B.}~\bibnamefont {Bauer}},\ and\ \bibinfo
  {author} {\bibfnamefont {A.~W.~W.}\ \bibnamefont {Ludwig}},\ }\href
  {https://doi.org/10.48550/ARXIV.2302.09094} {\bibinfo {title}
  {Measurement-induced entanglement transitions in quantum circuits of
  non-interacting fermions: Born-rule versus forced measurements}} (\bibinfo
  {year} {2023})\BibitemShut {NoStop}%
\bibitem [{\citenamefont {Yin}\ and\ \citenamefont
  {Barkai}(2022)}]{arrival_time_resetting}%
  \BibitemOpen
  \bibfield  {author} {\bibinfo {author} {\bibfnamefont {R.}~\bibnamefont
  {Yin}}\ and\ \bibinfo {author} {\bibfnamefont {E.}~\bibnamefont {Barkai}},\
  }\href {https://doi.org/10.48550/ARXIV.2205.01974} {\bibinfo {title} {Restart
  expedites quantum walk hitting times}} (\bibinfo {year} {2022}),\ \bibinfo
  {note} {arXiv:2205.01974}\BibitemShut {NoStop}%
\bibitem [{\citenamefont
  {Kurkij\"arvi}(1973)}]{random_barrier_heights_conductivity}%
  \BibitemOpen
  \bibfield  {author} {\bibinfo {author} {\bibfnamefont {J.}~\bibnamefont
  {Kurkij\"arvi}},\ }\bibfield  {title} {\bibinfo {title} {Hopping conductivity
  in one dimension},\ }\href {https://doi.org/10.1103/PhysRevB.8.922}
  {\bibfield  {journal} {\bibinfo  {journal} {Phys. Rev. B}\ }\textbf {\bibinfo
  {volume} {8}},\ \bibinfo {pages} {922} (\bibinfo {year} {1973})}\BibitemShut
  {NoStop}%
\bibitem [{\citenamefont {Machta}(1985)}]{qtm_asymptotics}%
  \BibitemOpen
  \bibfield  {author} {\bibinfo {author} {\bibfnamefont {J.}~\bibnamefont
  {Machta}},\ }\bibfield  {title} {\bibinfo {title} {Random walks on site
  disordered lattices},\ }\href {https://doi.org/10.1088/0305-4470/18/9/008}
  {\bibfield  {journal} {\bibinfo  {journal} {Journal of Physics A:
  Mathematical and General}\ }\textbf {\bibinfo {volume} {18}},\ \bibinfo
  {pages} {L531} (\bibinfo {year} {1985})}\BibitemShut {NoStop}%
\bibitem [{\citenamefont {Lawler}(1986)}]{random_walk_hitting_times}%
  \BibitemOpen
  \bibfield  {author} {\bibinfo {author} {\bibfnamefont {G.~F.}\ \bibnamefont
  {Lawler}},\ }\bibfield  {title} {\bibinfo {title} {Expected hitting times for
  a random walk on a connected graph},\ }\href
  {https://doi.org/https://doi.org/10.1016/0012-365X(86)90030-0} {\bibfield
  {journal} {\bibinfo  {journal} {Discrete Mathematics}\ }\textbf {\bibinfo
  {volume} {61}},\ \bibinfo {pages} {85} (\bibinfo {year} {1986})}\BibitemShut
  {NoStop}%
\bibitem [{\citenamefont {Szyniszewski}\ \emph {et~al.}(2022)\citenamefont
  {Szyniszewski}, \citenamefont {Lunt},\ and\ \citenamefont
  {Pal}}]{related_preprint}%
  \BibitemOpen
  \bibfield  {author} {\bibinfo {author} {\bibfnamefont {M.}~\bibnamefont
  {Szyniszewski}}, \bibinfo {author} {\bibfnamefont {O.}~\bibnamefont {Lunt}},\
  and\ \bibinfo {author} {\bibfnamefont {A.}~\bibnamefont {Pal}},\ }\href
  {https://doi.org/10.48550/ARXIV.2211.02534} {\bibinfo {title} {Disordered
  monitored free fermions}} (\bibinfo {year} {2022}),\ \bibinfo {note}
  {arXiv:2211.02534}\BibitemShut {NoStop}%
\bibitem [{\citenamefont {Bai}\ and\ \citenamefont
  {Breen}(2008)}]{com_periodic_bcs}%
  \BibitemOpen
  \bibfield  {author} {\bibinfo {author} {\bibfnamefont {L.}~\bibnamefont
  {Bai}}\ and\ \bibinfo {author} {\bibfnamefont {D.}~\bibnamefont {Breen}},\
  }\bibfield  {title} {\bibinfo {title} {Calculating center of mass in an
  unbounded 2d environment},\ }\href
  {https://doi.org/10.1080/2151237X.2008.10129266} {\bibfield  {journal}
  {\bibinfo  {journal} {Journal of Graphics Tools}\ }\textbf {\bibinfo {volume}
  {13}},\ \bibinfo {pages} {53} (\bibinfo {year} {2008})}\BibitemShut {NoStop}%
\end{thebibliography}%
\end{document}